\documentclass[10pt, peerreviewca, draftclsnofoot, onecolumn]{IEEEtran}

\usepackage[noadjust]{cite}
\usepackage[pdftex]{graphicx}
\usepackage{amsmath}

\usepackage{array}
\usepackage{fixltx2e}
\usepackage{url}
\usepackage{amsthm}

\usepackage[authoryear]{natbib}
\setcitestyle{open={(}, aysep{,}, authoryear, close={)}}

\usepackage{xcolor}
\usepackage{fancyhdr}

\usepackage{ulem}

\ifCLASSOPTIONcompsoc
\usepackage[caption=false,font=normalsize,labelfont=sf,textfont=sf]{subfig}
\else
\usepackage[caption=false,font=footnotesize]{subfig}
\fi

\theoremstyle{definition}
\newcommand{\njc}[1]{\textcolor{black}{#1}}

\begin{document}


\author{\IEEEauthorblockN{Nicholas Clark\IEEEauthorrefmark{1}, Matthew Dabkowski\IEEEauthorrefmark{2}, Patrick J. Driscoll\IEEEauthorrefmark{2}, Dereck Kennedy\IEEEauthorrefmark{2}, \\Ian Kloo\IEEEauthorrefmark{2}, Heidy Shi\IEEEauthorrefmark{1}}\\
\IEEEauthorblockA{\IEEEauthorrefmark{1}Department of Mathematical Sciences\\
\IEEEauthorrefmark{2}Department of Systems Engineering\\
U.S. Military Academy, West Point, New York 10996}
\\
Email: nicholas.clark@westpoint.edu\\
Ph: (+1) 845-938-1111  Fax: (+1) 845-938-1111}


\markboth{Pre-Print}{Clark \MakeLowercase{\textit{et al.}}: SUS}

\IEEEpubid{“This article has been accepted for publication in the \textit{International Journal of Human-Computer Interaction}, published by Taylor & Francis.”}



\IEEEpubidadjcol

\title{Empirical Decision Rules for Improving the Uncertainty Reporting of Small Sample System Usability Scale Scores}

\maketitle


\begin{abstract}
\textcolor{black}{The System Usability Scale (SUS) is a short, survey-based approach used to determine the usability of a system from an end user perspective once a prototype is available for assessment. Individual scores are gathered using a 10-question survey with the survey results reported in terms of central tendency (sample mean) as an estimate of the system's usability (the SUS study score), and confidence intervals on the sample mean are used to communicate uncertainty levels associated with this point estimate. When the number of individuals surveyed is large, the SUS study scores and accompanying confidence intervals relying upon the central limit theorem for support are appropriate. However, when only a small number of users are surveyed, reliance on the central limit theorem falls short, resulting in confidence intervals that suffer from parameter bound violations and interval widths that confound mappings to adjective and other constructed scales. These shortcomings are especially pronounced when the underlying SUS score data is skewed, as it is in many instances. This paper introduces an empirically-based remedy for such small-sample circumstances, proposing a set of decision rules that leverage either an extended bias-corrected accelerated (BCa) bootstrap confidence interval or an empirical Bayesian credibility interval about the sample mean to restore and bolster subsequent confidence interval accuracy. Data from historical SUS assessments are used to highlight shortfalls in current practices and to demonstrate the improvements these alternate approaches offer while remaining statistically defensible. A freely available, online application is introduced and discussed that automates SUS analysis under these decision rules, thereby assisting usability practitioners in adopting the advocated approaches.}
\end{abstract}
\IEEEpeerreviewmaketitle


\section*{Disclaimer}
\addcontentsline{toc}{section}{Disclaimer}
The views expressed herein are those of the authors and do not reflect the position of the United States Military Academy, the Department of the Army, or the U.S. Department of Defense.

\section{Introduction}\label{s.introduction}

\IEEEPARstart{T}he System Usability Scale (SUS) has been employed for over 20 years as a reliable end-of-test subjective assessment tool to evaluate the perceived usability of a system \citep{brooke2013sus}. First introduced by \cite{brooke1996sus}, it has been used extensively in industry to provide valid and consistent design feedback for assessing the usability of human-machine systems, software, and websites \citep{peres2013validation}, as well as everyday products \citep{kortum2013usability}. Usability in these settings encompass a broader scope than ``ease of use'' or ``user friendliness'' \citep{ISO9241}, dimensions of usability that affect system technical acceptance by end users \citep{King2006}. Among the SUS's greatest strengths is the simplicity of its design. As seen in Table 1, the SUS is a Likert scale-based survey composed of ten questions with five levels of response used to elicit a user’s level of agreement concerning system characteristics. 

\begin{table}[ht]
\begin{center}
\begin{tabular}{| p{0.5cm} | m{10cm} |}
\hline
Q1 & I think that I would like to use this system frequently. \\ 
 Q2 & I found the system unnecessarily complex. \\ 
 Q3 & I thought the system was easy to use. \\ 
 Q4 & I think that I would need the support of a technical person to be able to use this system. \\ 
 Q5 & I found the various functions in this system were well integrated. \\ 
 Q6 & I thought there was too much inconsistency in this system. \\ 
 Q7 & I would imagine that most people would learn to use this system very quickly. \\ 
 Q8 & I found the system very cumbersome to use. \\ 
 Q9 & I felt very confident using the system. \\ 
 Q10 & I needed to learn a lot of things before I could get going with this system. \\ 
 \hline
\end{tabular}
 \vspace{0.25cm}
\caption{Standard SUS questions.}
\textsuperscript{a}Respondents indicate their agreement with each statement from ``1 = Strongly Agree" to ``5 = Strongly Disagree."
\label{sample-table}
\end{center}
\end{table}

Post-survey analysis regarding individual SUS questions typically proceeds by treating responses as interval data and using appropriate parametric analysis \citep{likertexplored, boone2012analyzing, carifio2008resolving}. Of primary interest in systems analysis is the aggregate of each respondent's answers, dubbed a \textit{SUS score}. This SUS score estimates an individual's subjective judgment regarding the usability of a system. The SUS score is obtained by converting each question's response $x_{i} \in \{1, 2, 3, 4, 5\}$, for $i=1,\dots,10$, into a single value depending on whether the question was positively worded (the odd numbered questions): $(x_i - 1)$, or negatively worded (the even numbered questions): $(5 - x_i)$, and then multiplying the sum of these by 2.5, thus bounding a respondent's SUS score to the interval [0, 100] in increments of 2.5 units. The sample average of SUS scores across all $n$ SUS respondents yields a \textit{SUS study score}. The overall SUS results for a system are often reported as a single SUS study score \citep{koltringer2004comparing, lewis2018item, tullis2004comparison}, a SUS study score with a standard error \citep{everett2006measuring, kortum2015measuring, bangor2009determining}, or a standard confidence interval (CI) for the mean of the SUS study score constructed from a \textit{z} or \textit{t} statistic \citep{blavzica2015slovene, borsci2015assessing, orfanou2015perceived}.

 
\begin{figure}
\centering
\includegraphics[width=1\linewidth]{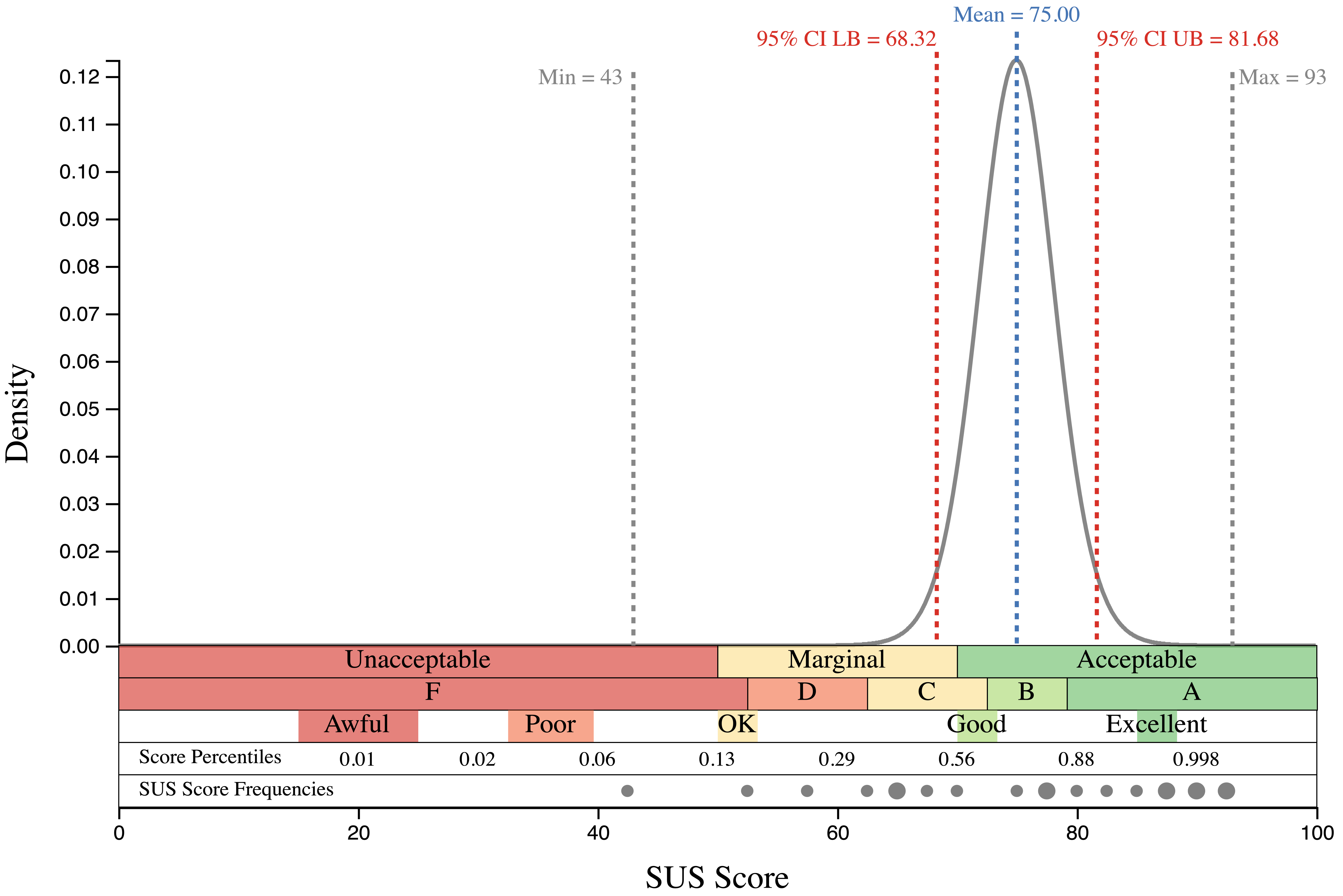}
\caption{SUS scores mapped to published usability scales, namely: acceptability ranges (from Figure 13 of Bangor et al. \citeyearpar[p.~592]{bangor2008empirical}), letter grades (from Table 8.5 of Sauro and Lewis \citeyearpar[p.~204]{sauro2016quantifying}), adjective ratings with 95\% CIs (derived from Table 3 of Bangor et al. \citeyearpar[p.~118]{bangor2009determining}), and score percentiles (from Table 8.4 of Sauro and Lewis \citeyearpar[p.~203]{sauro2016quantifying}).}
\label{f:SUSscoreMapping1}
\end{figure}


In some applications, organizations establish specific descriptors associated with numerical results based on past experience with similar systems. When organizations map SUS study scores to pre-defined usability labels in this manner using acceptability ranges \citep{bangor2008empirical}, letter grades \citep{sauro2016quantifying}, adjective ratings \citep{bangor2009determining}, or score percentiles \citep{sauro2016quantifying}, SUS study scores can be communicated to project managers and acquisition officials in an intuitive and actionable way. Figure~\ref{f:SUSscoreMapping1} shows an example mapping of SUS scores to all four of the noted usability labels. Here, all four mappings are shown for convenience in exposition as typically an organization would adopt only one for their use. 

Mapping the CI bounds of the SUS study score to usability labels adds further insights into survey results by portraying underlying uncertainty associated with SUS study results. When the CI for the SUS study score is mostly contained in a single usability label's interval, sufficient evidence supports the usability characterization associated with that label. For example, Figure~\ref{f:SUSscoreMapping1} SUS results support the associated system's usability being labeled as nearly 100\% acceptable with little data in the category labeled `Marginal'. On the other hand, CIs that cross interval bounds can complicate one's usability interpretation. Returning to Figure~\ref{f:SUSscoreMapping1}, mapping the same CI to letter grades indicates that the system's usability ranges between a high C and low A, adjective ratings indicate the results existing mostly in the category labeled `Good' with high side spillover into `Excellent', and Score Percentiles are concentrated between roughly 50 and 90\%  when compared to previously evaluated similar systems. One can see from this example that tighter CIs are more desirable when SUS results are used in this context because they narrow the range of mapped SUS score interpretations. 

The SUS has undergone extensive psychometric evaluation since its introduction, consistently demonstrating its reliability \citep{Lucey1991,bangor2008empirical,lewissauro2009,sauro2016quantifying} and content validity \citep{finstad2010, lewisplus2013}. Moreover, these favorable characteristics appear to persist despite language translations \citep{dianat2014psychometric, blavzica2015slovene, katsanos2012perceived}, using English language versions with non-native English speakers \citep{finstad2006}, modifications that use an all-positive version of the survey \citep{sauro2011designing}, as well as removing a single test item from the questionnaire \citep{lewis2017can} if deemed inappropriate for use with the system at-hand. In this last instance, practitioners must adjust the multiplier from 2.5 to 2.78 to accommodate the reduction in test items from 10 to 9. 

SUS scores have been shown to be sensitive to types of interfaces and changes to a product \citep{bangor2008empirical}, the task order used for assessment \citep{tullis2004comparison}, user experience \citep{kortum2013usability, mclellan2012effect}, and differences in user age but not gender \citep{bangor2008empirical}.  Moreover, the SUS has been evaluated for efficacy against other known methods for testing usability such as the Computer System Usability Questionnaire (CSUQ) and the Usability Metric for User Experience (UMUX) \citep{lewis2018measuring, tullis2004comparison, borsci2015assessing}. Interest in capturing, confirming, and expounding on SUS utility through research continues to grow. Brooke's \citeyearpar{brooke1996sus} seminal article on the SUS reported 8,948 Google Scholar citations as of 20 May 2020, a gain of 3,284 since reported by \cite{lewis2018past} two years earlier.



While the vast majority of research efforts have focused on validating the SUS as a sound methodology and leveraging its normative data to evaluate system usability, there remains opportunity to improve the analysis and reporting of SUS study scores when practitioners are faced with very small sample ($n \leq 10$) and extremely small sample ($n \leq 6$) circumstances. In practice, this most often happens when cost constraints, security considerations, system operational complexity, or the limited availability of users with highly specialized skills impose on the system assessment process \citep{chen1995}. 

When the system under assessment represents a capability gap-filling, new technology prototype, it tends to also receive more enthusiastically positive responses from users who quickly comprehend the system's potential to solve or partially-solve nagging operational deficiencies. This effect, along with its counterpart, introduces skewness in the data that further complicates statistical methods. The challenge is exacerbated still when assumptions regarding the continuity of the underlying distribution are not appropriate. In this context, we have two concerns.  

Firstly, the distribution of historical SUS study scores is skewed \citep{bangor2008empirical, sauro2016quantifying}, violating symmetry assumptions that call into question methods relying on the central limit theorem (CLT) when the sample size is small. Historically, SUS study scores across all usability studies \textcolor{black}{analyzed for this paper} have an average skewness of about -0.4  \citep{lewis2009factor}, as given in \cite{sauro2016quantifying}. Data from \cite{bangor2008empirical} demonstrate that for a \textcolor{black}{single} study, this skewness can range from highly negative to highly positive. \textcolor{black}{One logical explanation for this characteristic is that} by the time users engage with system prototypes, features are refined to a point that a majority of users will find the system under investigation to be reasonably simple to use. 

\begin{figure}[h]
\centering
\includegraphics[width=1\linewidth]{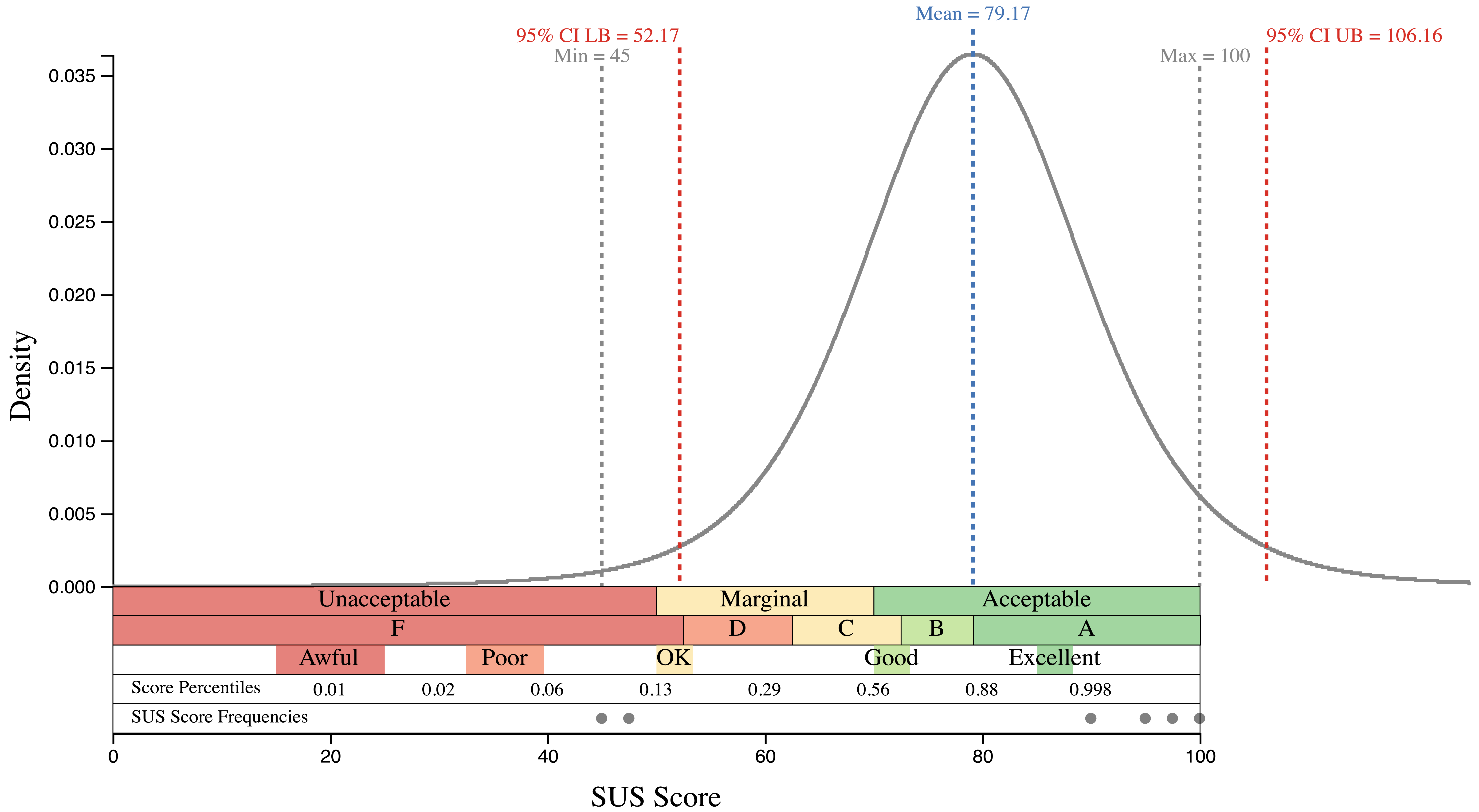}
\caption{Upper CI bound for the mean SUS score violating the upper bound of the parameter space.}
\label{f:SUSscoreBoundViolation1}
\end{figure}

Secondly, SUS study scores are bounded to the interval $[0, 100]$ in discrete increments of 2.5.  Therefore, a system perceived as highly usable will tend to produce SUS scores closer to the upper bound. The theory allowing practitioners to use standard methods for constructing confidence intervals relies on asymptotics that manifest slowly when skewness is present, a condition exacerbated when skewness is high. The resulting negative skewness in SUS studies causes the upper bounds of CIs to violate the parameter space of the scoring interval in small sample studies where higher levels of variability are expected. Figure~\ref{f:SUSscoreBoundViolation1} shows an example of this for a SUS study with 6 participants, where 4 of the 6 SUS scores were near the upper bound of 100, yielding a sample skewness of -0.9.

From a practical perspective, if small sample SUS studies were uncommon, the concerns identified would be real but somewhat irrelevant. This is not the case. In particular, determining the usability of highly complicated or complex systems tends to be restricted to small sample sizes, mainly due to limited accessibility to these systems and cost considerations involving highly specialized test subjects. As some datasets suggest (e.g., \cite{bangor2008empirical}), very small and extremely small SUS studies are effectively the rule rather than the exception, and in such situations, reporting a standard error or using traditional confidence intervals neglects potential statistical issues that exist in the data. Ultimately, the method of CI construction should generate narrow confidence intervals with nominal coverage of the underlying population mean.

In this paper, the impact of very small sample ($n \leq 10$) SUS study results are examined with specific focus on achieving desirable characteristics for a reported SUS study confidence interval, and the accessibility of potential SUS study scores given the discrete nature of the underlying survey scale for individual responses. The results are intended to assist usability professionals in increasing the accuracy and validity of their SUS results when faced with a very small number of SUS respondents.

In Section~\ref{s:methods}, two alternatives to the common practice of using \textit{t} distribution-based confidence intervals for small sample SUS study scores are examined to assess their efficacy towards achieving the previously stated desirable characteristics. Section~\ref{s:simulation_study} applies these alternative approaches to a repository of data from actual SUS studies used with permission from the owners. Results demonstrate achievable improvements in reporting accuracy regarding SUS study results. In Section~\ref{s:app}, a publicly available, custom R-based application that implements the three major computational approaches presented in this paper is described. This application is intended to enable practitioners to leverage the recommendations provided herein in an efficient manner. Section~\ref{s:conclusions} summarizes this paper's major results and recommendations along with potential future opportunities. 

\section{Methods for Quantifying Uncertainty}\label{s:methods}

Generally speaking, the CLT states that the distribution of the sample mean approximates a normal distribution as the sample size becomes large, regardless of the shape of the population's distribution. Oftentimes, sample sizes greater than 30 are considered sufficient for the CLT to hold. In such cases, a CI that reflects a range of plausible values for the population mean ($\mu$) can be calculated using the standard expression: 
\[{\bar{x}} \pm z_{{{\alpha}/2}} \frac{s}{\sqrt{n}}\]

\noindent where ${\bar{x}}$ is the sample mean, $z_{{\alpha}/2}$ is the critical value from the standard normal distribution evaluated at ${\alpha}/2$ when $1-{\alpha}$ is the confidence level, $s$ is the sample standard deviation, and $n$ is the sample size.

\begin{figure}
\centering
\includegraphics[width=0.8\linewidth]{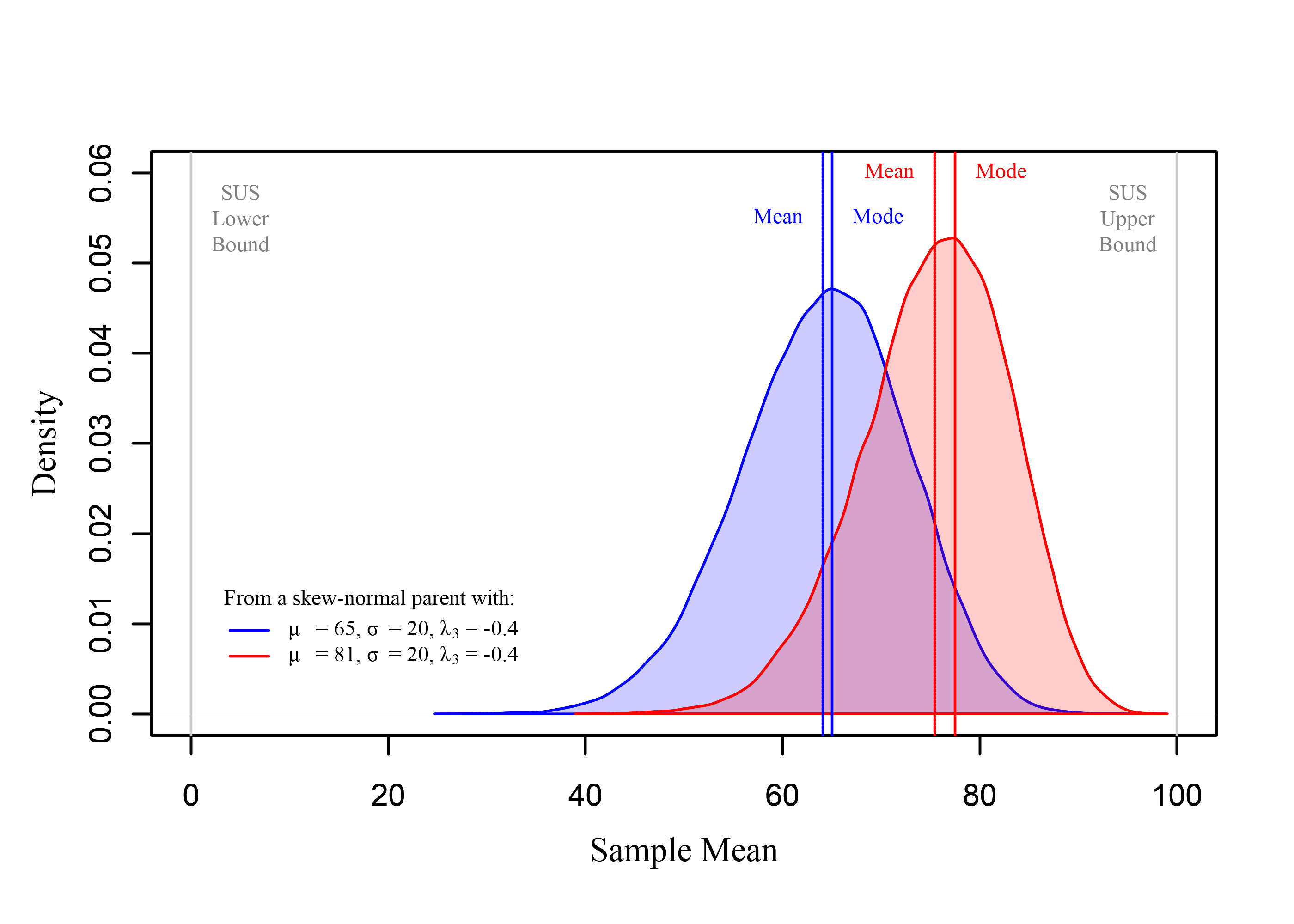}
\caption{Density plots of the sample means from 100,000 samples of size $n=5$ from skew-normal distributions.}
\label{fig:simmeans}
\end{figure}

For the reasons mentioned earlier, the distribution of the sample mean $\bar{X}$ is likely not symmetric for small samples. To demonstrate this, assume that the underlying distribution of SUS scores follows Azzalini's \citeyearpar{azzalini2005skew} skew-normal distribution with a population mean ($\mu$), standard deviation ($\sigma$), and skewness ($\lambda_3$) of 65, 20, and -0.4, respectively. Due to the upper truncation of the distribution at 100, these correspond to $\mu = 63$, $\sigma = 19$, and $\lambda_3 = -0.45$, which closely mirror the values reported in Lewis and Sauro \citeyearpar{lewis2009factor}. To visualize the distribution of the sample mean when $n=5$, 100,000 random samples from this skew-normal parent distribution were generated with realizations rounded to the nearest 2.5 to match the domain of observable SUS scores. As seen in the blue density plot shown in Figure~\ref{fig:simmeans}, the sample mode sits to the right of the sample mean and the skewness of $\bar{X}$ is approximately -0.22. Moreover, as the population mean gets closer to the SUS score's upper bound, the magnitude of $\bar{X}$'s skewness increases. For instance, if the mean shifts to 81, the skewness of $\bar{X}$ becomes roughly -0.38, which is noticeable in Figure~\ref{fig:simmeans}'s red density plot. While rules of thumb suggest that skewness values between -1 and 1 are acceptable, normal-based confidence intervals often fail to achieve nominal coverage within this range.

To understand why the CLT may be inadequate for very small sample SUS studies, it is beneficial to consider the Edgeworth expansion for the studentized sample mean: $n^{1/2} \frac{(\bar{X}-\mu)}{\sigma}$.  As given in \cite{hall2013bootstrap}, the probability density function (PDF) of this statistic can be expressed by:

\begin{equation}
    f(x) = \phi(x)-\frac{1}{\sqrt{n}}\frac{\lambda_3}{6}\phi ^{(3)}(x) + O(n^{-1})\label{eq:edgeworth}
\end{equation}

\noindent where $\phi^{(i)}(x)$ is the $i$th derivative of the standard normal distribution and $\lambda_3$ is the third central moment (i.e., skewness) over the standard deviation cubed.  Therefore, with skewness present the distribution of the sample mean approaches a standard normal at a rate of $n^{-1/2}$, whereas in the absence of skewness it approaches a standard normal at a rate of at most $n^{-1}$. Practically, this means that if $n$ is small (i.e., few respondents participating in a SUS study), the distribution of the sample mean should still have some amount of skewness present. When sample sizes are small, the standard normal distribution is often replaced with the Student's \textit{t} distribution to generate the critical value to construct the CI:
\[{\bar{x}} \pm t_{{{\alpha}/2,n-1}} \frac{s}{\sqrt{n}}\]

\noindent where $t_{{{\alpha}/2,n-1}}$ corresponds to the $\alpha/2$ quantile from a $t$ distribution with $n-1$ degrees of freedom. In such situations, a symmetric CI such as one formed using the \textit{t} distribution, may be inappropriate, with this condition becoming exacerbated as $n$ approaches one. On the other hand, when $n$ is large, the $\frac{1}{\sqrt{n}}$ term in \eqref{eq:edgeworth} goes to zero, meaning the distribution of the sample mean becomes symmetric.

\subsection{Options for Confidence Interval Formulation}\label{ss.options}

Despite the lack of evidence supporting a CLT assumption for small sample data, it does not necessarily mean that using the CLT is categorically bad or always inappropriate in such cases. Several researchers have proposed modifications to the $t$ distribution approach when sample sizes are as small as 13 \citep{johnson1978, sutton1993}, although as researchers note, results can be quite inaccurate. And, although the \textit{t} distribution has nice coverage properties, it is not guaranteed to obey the parameter space. For example, consider the case when $n = 5$. Suppose that among the five SUS responses, three respondents find the product exceptional, and two rate it as good, resulting in potential SUS scores of \njc{97.5, 97.5, 97.5, 80, and 80}. In this case, the 95\% CI formed by using a \textit{t} distribution is \njc{(78.5, 102.4)}, which is nonsensical as $\mu$ cannot be greater than 100.  As described earlier, this occurs because using the \textit{t} distribution  assumes that the distribution of $\bar{X}$ is symmetric, and when $n$ is small this is not guaranteed to be true. 

\textcolor{black}{While it is tempting to truncate the above CI to (78.5, 100), the result is no longer a valid 95\% confidence interval. In general, a confidence interval of a parameter, $\mu$, is of level $1-\alpha$ if $P(L(X)<\mu<U(X))=1-\alpha$ \citep{casella2002statistical}. Using the \textit{t} distribution in the above example yields a 95\% CI with $L(x) = 78.5$ and $U(x) =102.4$. Inherent to this construction is the belief that $P(\bar{X} \in (100,102.4)|\mu)>0$, and in this case, the confidence interval calculation used implies $P(\bar{X} \in (100,102.4)|\mu)\approx 0.045$. Simply truncating the interval at 100 renders the probability that $\bar{X}$ is above 100 to be zero, resulting in a confidence interval with less than its nominal coverage \citep{agresti1998approximate}, perhaps significantly \citep{mandelkern2002setting}. Under these conditions, a practitioner should not be comfortable reporting results at the nominal level. In order to guarantee a 95\% confidence interval in this instance, the lower bound must be shifted left to account for the missing 4.5\% probability. In other words, a properly constructed CI requires a   $L_{\mbox{shifted}}(X)$ such that $P(L_{\mbox{shifted}}(X)<\mu<100)=1-\alpha$. In the above example, this would result in an interval of (70, 100). Depending on the adjective label intervals being used, the usability of a system might easily be considered marginal instead of acceptable.}

\textcolor{black}{An additional detractor to a truncating approach is that the distribution for $\bar{X}$ is no longer valid. To fix this, one might abandon the belief that $\bar{X}$ follows a non-central \textit{t} distribution and instead assume that it follows a truncated \textit{t} distribution, and hence formulate a one-sided confidence interval. While explorations involving the truncated \textit{t} distribution are beyond the scope of this paper, preliminary simulations suggest that using this distribution may have benefits. Additional alternate ways to form a confidence interval on a bounded parameter space are given in Wu and Neale \citeyearpar{wu2012adjusted}, Bebu and Mathew \citeyearpar{bebu2009confidence}, and Andrews \citeyearpar{andrews1999estimation}.}

In sum, while the \textit{t} distribution may validly address some issues with small samples, it fails to account for bounds on the underlying sample space for $\mu$ in SUS applications, and it does not account for potential skewness in the distribution of the sample mean. As demonstrated in  Section~\ref{s:simulation_study} that follows, many SUS studies limited to small sample sizes encounter this issue \citep{bangor2008empirical}. Fortunately, two alternatives offer relief: an expanded version of the bias-corrected accelerated bootstrap approach \citep{efron1987better} and uncertainty bounds created using a Bayesian credible interval. 

\subsection{CIs Built Using the Expanded Bias-Corrected Accelerated Bootstrap}\label{ss.expandedBCa}

In general, bootstrap methods are empirical statistical sampling approaches used to estimate characteristics of unknown distributions when faced with small sample sizes or inappropriate use of parametric assumptions. In parametric bootstrap methods, random samples are generated from a parametric model fit to the data. In non-parametric resampling, bootstrap samples are constructed using resampling with replacement from the original sample \citep{EfronTib1986, Kysely2010}. Given the limitations noted earlier,  a non-parametric random sampling approach is warranted for this exploration \citep{diciccio1988, Flowers2018}.

In comparison to the detractors present in a \textit{t} distribution approach noted earlier, a percentile bootstrap CI for the population mean ($\mu$) is simple to form, easy to understand, and guaranteed to obey bounds on a parameter space. It uses the ${100({\alpha/2})}$ and the ${100(1-{\alpha/2})}$ percentiles of the bootstrap distribution as its bounds, where ${\alpha}$ is the likelihood that $\mu$ lies outside of the CI. For example, to construct a 95\% CI using the percentile bootstrap method, ${\alpha} = 0.05$, and the CI would be constructed using the 2.5 and 97.5 percentiles of the bootstrap distribution. 

To generate sufficient realizations of a SUS sample mean, the percentile bootstrap CI for $\mu$ resamples from SUS score data with replacement $B$ times to form $B$ bootstrap samples. At each iteration, the sample mean is calculated for each bootstrap sample. Taken together, these $B$ realizations of the sample mean approximate a sampling distribution for the population mean by ordering them from smallest to largest such that: 
\[{\hat{\theta^*}_{(1)}} \leq {\hat{\theta^*}_{(2)}} \leq {\hat{\theta^*}_{(3)}} \leq ... \leq {\hat{\theta^*}_{(B)}},\]
where ${\hat{\theta^*}_{(i)}}$ represents $i^{th}$ smallest sample mean for $i= 1, 2, \ldots, B$. Using this approach, the corresponding 95\% confidence interval for the population mean is given by:
\[[{\hat{\theta^*}_{(0.025B)}},{\hat{\theta^*}_{(0.975B)}}].\]

\noindent As an illustration, consider the small SUS score dataset introduced earlier, namely \{97.5, 97.5, 97.5, 80, 80\}, and set $B = 1,000$. The percentile bootstrap with $B = 1,000$ resamples 1,000 datasets with $n = 5$, subsequently calculating the sample mean for each dataset. With  resampling with replacement, it is possible to obtain bootstrap samples with many repeated values, such as \{80, 80, 80, 80, 80\}, which would result in a mean of 80. Once  resampling is complete and 1,000 sample means are calculated, the 2.5\textsuperscript{th} and 97.5\textsuperscript{th} smallest values would be extracted and used as the 95\% confidence interval's lower and upper bounds, respectively.

To account for potential skewness in data, \cite{efron1987better} introduced a bias-corrected and accelerated (BCa) bootstrap methodology for both parametric and nonparametric situations, improving an earlier percentile method \citep{efron1981}. Similar to confidence intervals constructed using the percentile bootstrap method, bias-corrected and accelerated (BCa) bootstrap CIs are formed using percentiles of the bootstrap distribution. However, unlike the percentile bootstrap method, the choice of which percentiles to use is more complicated and requires one to calculate two additional factors: a bias correction factor $b$ and an acceleration factor $a$. The bias correction factor $b$ is calculated as $b=\Phi^{-1}(p)$, where $\Phi$ is the cumulative distribution function of a standard normal random variable and $p$ is the proportion of bootstrap samples less than the average.  This factor estimates the difference between the median of the bootstrap distribution and $\bar{X}$.  Clearly, if there is no skew to the sampling distribution of $\bar{X}$, this bias correction factor will be zero on average. 

The acceleration factor $a$ is obtained by jackknife resampling of the $n$ original data to estimate the second term in \eqref{eq:edgeworth} using the following relation: 

\begin{equation}
    a= \frac{1}{\sqrt{n}}\frac{\lambda_3}{6}\phi ^{(3)}(x).
\end{equation}

\noindent This involves generating $B$ replicates of the original sample. The first jackknife sample leaves out the first value of the original sample, the second by leaving out the second value, and so on until $B$ samples of size $(n-1)$ are obtained \citep{Carpenter2000}. Once this term is estimated, both the bias correction and the acceleration terms can be used to make a bootstrap estimate that converges at a rate of $O(n^{-1})$ rather than $O(n^{-1/2})$ as shown in \cite{hall2013bootstrap}. 

Although the BCa bootstrap method is an improvement over the percentile bootstrap technique, the CIs produced by the BCa bootstrap method tend to be too narrow for small samples and may fail to achieve their nominal coverage probability. In other words, when the sample size is small, BCa bootstrap CIs may not cover their parameters' true values as often as they claim to. One approach to addressing this issue is given in \cite{hesterberg2015teachers} and adopted herein, which is similar to using a \textit{t} distribution instead of the CLT-based normal distribution to form traditional CIs. Using a \textit{t} distribution instead of a normal distribution is the same as multiplying the length of a CI by $\displaystyle (s \times t_{\alpha/2,n-1})/(\hat{\sigma} \times z_{\alpha/2})$. If the underlying distribution is not normally distributed, applying this correction is not theoretically sound, but it is a commonly used correction factor in practice. 

\begin{figure}
\centering
\includegraphics[width=1\linewidth]{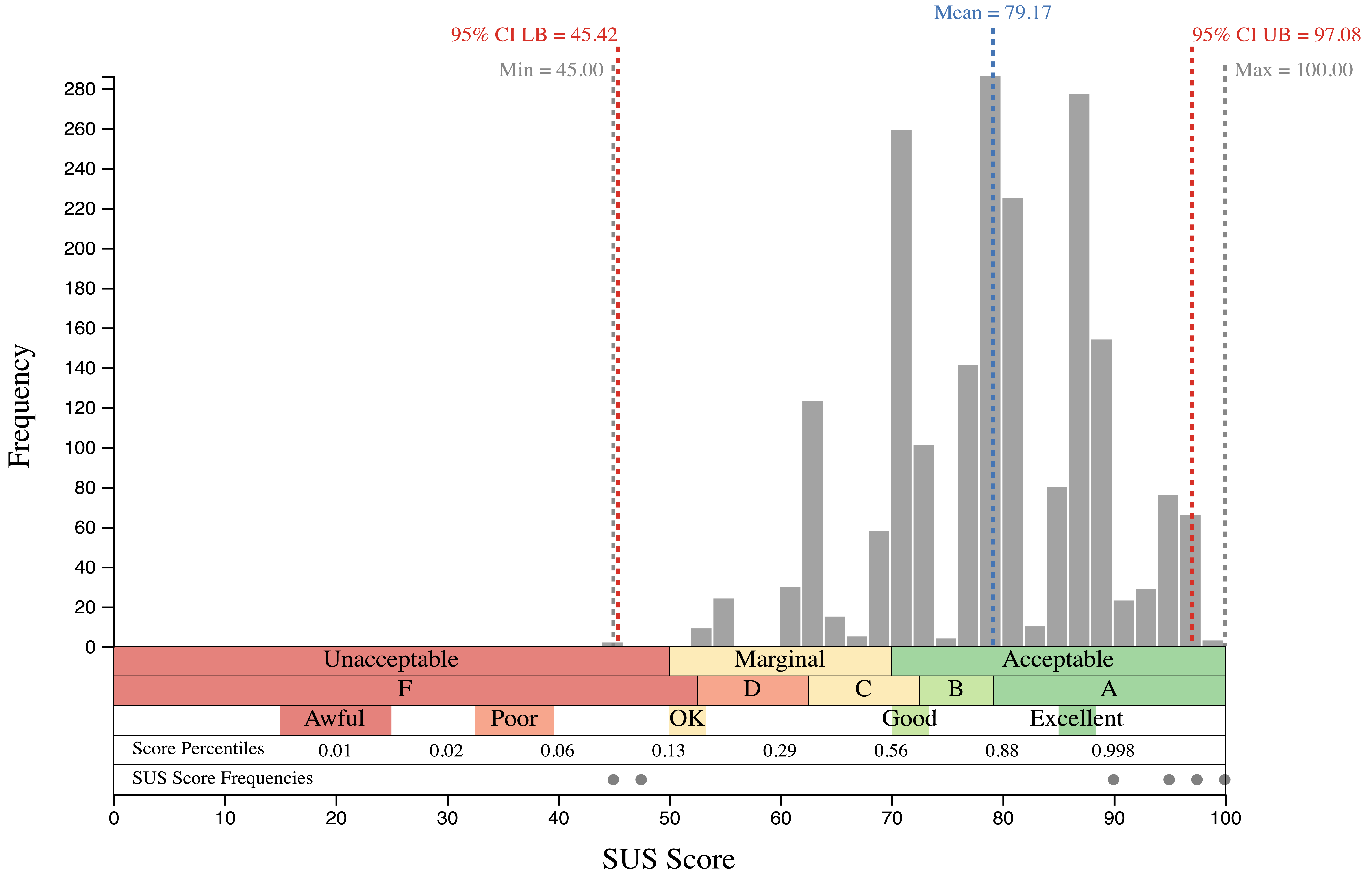}
\caption{95\% expanded BCa bootstrap CI for the mean SUS score of the survey highlighted in Figure~\ref{f:SUSscoreBoundViolation1}.}
\label{f:SUSscoreBCa}
\end{figure}

This same approach can be applied to bootstrap CIs in a straightforward manner, most easily explained using the percentile bootstrap as an example. The percentile bootstrap uses the $100(\alpha/2)$ and the $100(1-\alpha/2)$ highest values from the bootstrap sample to form a $100(1-\alpha)\%$ CI. The expanded percentile bootstrap sets $\alpha'/2= \Phi (-\sqrt{\frac{n}{(n-1)}} t_{\alpha/2,n-1})$, where $\Phi$ is the standard normal CDF and $t_{\alpha/2,n-1}$ is the critical value found from the t distribution. The expanded bootstrap then uses the $100(\alpha'/2)$ and the $100(1-\alpha'/2)$ values from the bootstrap sample. This expansion can also be easily applied to the BCa bootstrap using the \texttt{resample} library \citep{resample} within the statistical software R. \njc{When applied to the example dataset introduced earlier: \{97.5, 97.5, 97.5, 80, 80\}, the expanded BCa bootstrap yields an interval of (80, 97.5) and thereby supporting the conclusion that the system's usability is acceptable.} Additionally, Figure~\ref{f:SUSscoreBCa} illustrates a 95\% expanded BCa bootstrap CI for the same SUS study highlighted in Figure~\ref{f:SUSscoreBoundViolation1}. 

Comparing the \textit{t} distribution CI in  Figure~\ref{f:SUSscoreBoundViolation1} to the expanded BCa bootstrap CI in Figure~\ref{f:SUSscoreBCa} reveals two important differences. First, unlike the upper bound of the CI derived from the \textit{t} distribution (106.16), the upper bound of the BCa bootstrap CI (97.50) abides the parameter space of the SUS score. In essence, a CI represents a plausible range of values for a parameter of interest. CIs that exclusively cover the feasible values for this parameter should be preferred. In the case of mean SUS scores, values below 0 or above 100 are not realizable, and hence not feasible. Additionally, the width of the expanded BCa bootstrap CI (51.67) is slightly narrower than the \textit{t} distribution CI (53.99). In general, when choosing between multiple valid CIs, the narrowest interval is preferred, assuming that the CI's construction method preserves the nominal coverage probability. While in this instance the practical conclusions would not change if an analyst selected the expanded BCa CI over the \textit{t} distribution CI, in some instances they would. For example, consider the small dataset presented earlier (i.e., \{97.5, 97.5, 97.5, 80, 80\}). After proper truncation the adjusted lower bound of the \textit{t} distribution CI would lead a practitioner to conclude that the system's usability is unacceptable, whereas the expanded BCa CI would not.

The above discussion indicates the expanded BCa bootstrap CI is the better option for this small sample SUS study, and Section \ref{s:simulation_study} investigates the generalizability of these results using simulation. However, before moving on, there is a final, more subtle point worth mentioning. 

\begin{figure}[h]
\centering
\includegraphics[width=0.6\linewidth]{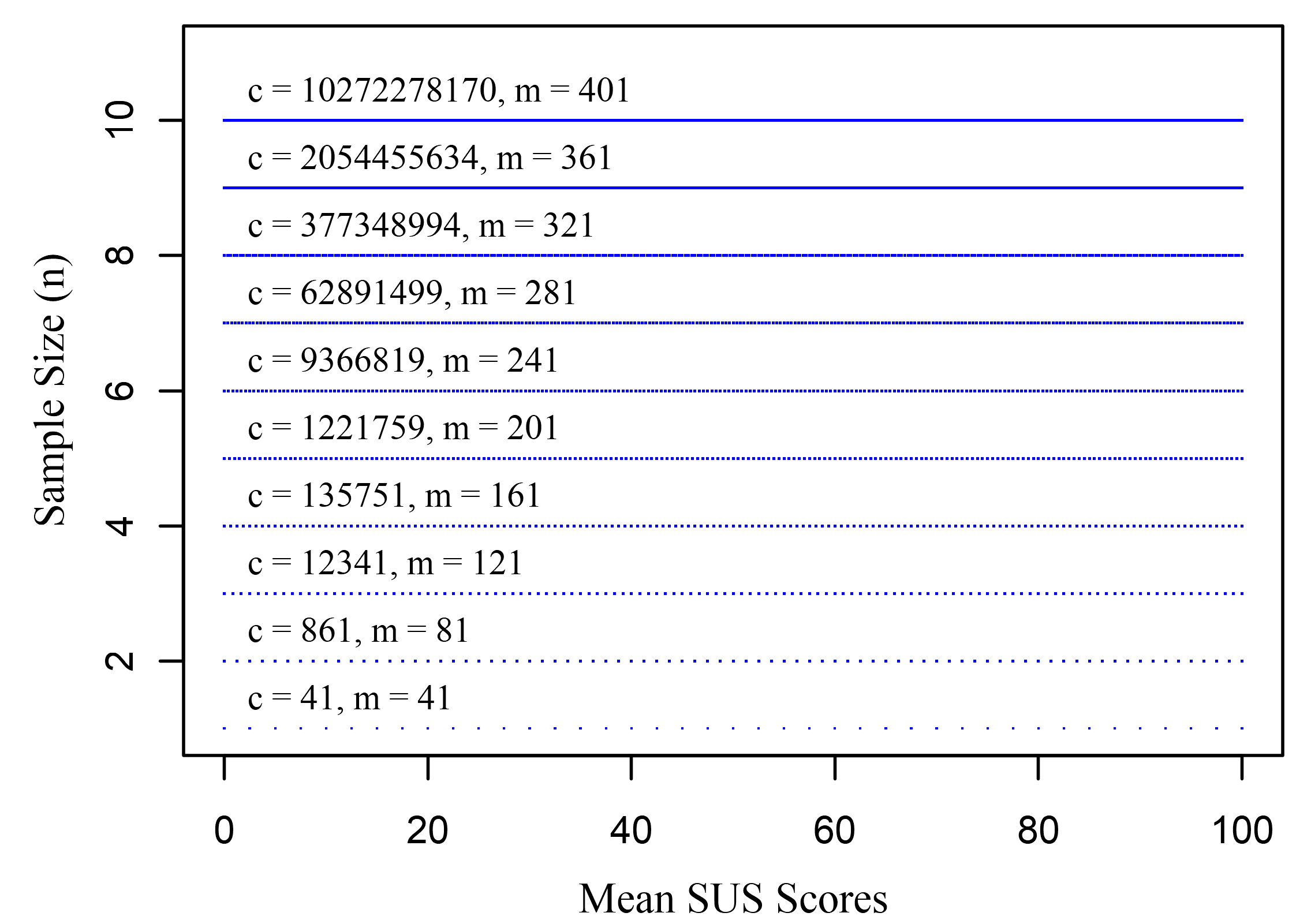}
\caption{Possible mean SUS scores for samples of size \textit{n}, where \textit{c} is the number SUS score combinations of size \textit{n} and \textit{m} is the number of distinct mean SUS scores.}
\label{f:DistinctSUSMeans}
\end{figure}

When confronted with small samples, using the \textit{t} distribution assumes the distribution of the underlying data is normally distributed or at least reasonably symmetric and continuous. As noted earlier, SUS scores often lack symmetry; case in point, the sample skewness of the SUS scores in Figure~\ref{f:SUSscoreBoundViolation1} is -0.9. Moreover, while SUS scores are clearly discrete, leading SUS literature currently suggests the sampling distribution of the mean SUS score is effectively continuous, stating ``the combination of average SUS scores for a study is virtually infinite” \citep[p. 202]{sauro2016quantifying}. Figure~\ref{f:DistinctSUSMeans} suggests otherwise, especially for small \textit{n}.  

As seen in Figure~\ref{f:DistinctSUSMeans}, when $n=6$ over 9 million SUS score combinations are possible, but there are only 241 distinct SUS means available. Increasing the sample size to $n=10$ produces more than 10 billion SUS score combinations. However, these combinations yield a mere 401 distinct SUS means. This is far from infinite, and the counter-intuitive result is a function of the SUS scores' special structure. Specifically, prior to scaling an individual SUS score by 2.5, the 41 possible scores are $S=\{0,1,2,...,40\}$, which can be rewritten as $S=\{s+id:i=0,1,...,k-1\}$ with $s=0$, $d=1$, and $k=41$. With $s \in S$ and $d \geq 1$, $S$ is a 41-term arithmetic progression. Leveraging a result from number theory, the mean SUS score for a sample of size \textit{n} is simply an n-fold \textit{sumset} of $S$ (scaled by $2.5/n$), and when $n \geq 2$ this sumset's cardinality is $41n-(n-1)$ \citep[p. 335]{mistri2014generalization}. Each time the sample size increases 1, the number of distinct SUS means increases by 40. When \textit{n} is small, these means are sparsely distributed along the real line between 0 and 100. Accordingly, treating the SUS mean as continuous and using a CI construction method that relies on this continuity, such as the \textit{t} distribution, appears ill-advised. This sentiment is reinforced in recent literature, notably \citet{liddell2018analyzing}. Fortunately, the expanded BCa bootstrap CI accommodates discrete data and provides a defensible alternative CI construction both in theory and practice.

\subsection{Credible Intervals Constructed Using an Empirical Bayesian Approach}\label{ss.empiricalBayes}

Although the expanded BCa bootstrap method is an attractive option as it obeys the parameter space for $\mu$, it does ignore any preconceived belief or historical evidence regarding a given system's mean SUS score. For example, it fails to account for the extremely low probability that the population mean is 0 or 100. Furthermore, as we show in Section~\ref{s:simulation_study}, when $n$ is less than 5, CIs formed using even the expanded BCa bootstrap method do not cover the true mean as often as they purport to do. 

\begin{figure}[!t]
\centering
\includegraphics[width=0.65\linewidth]{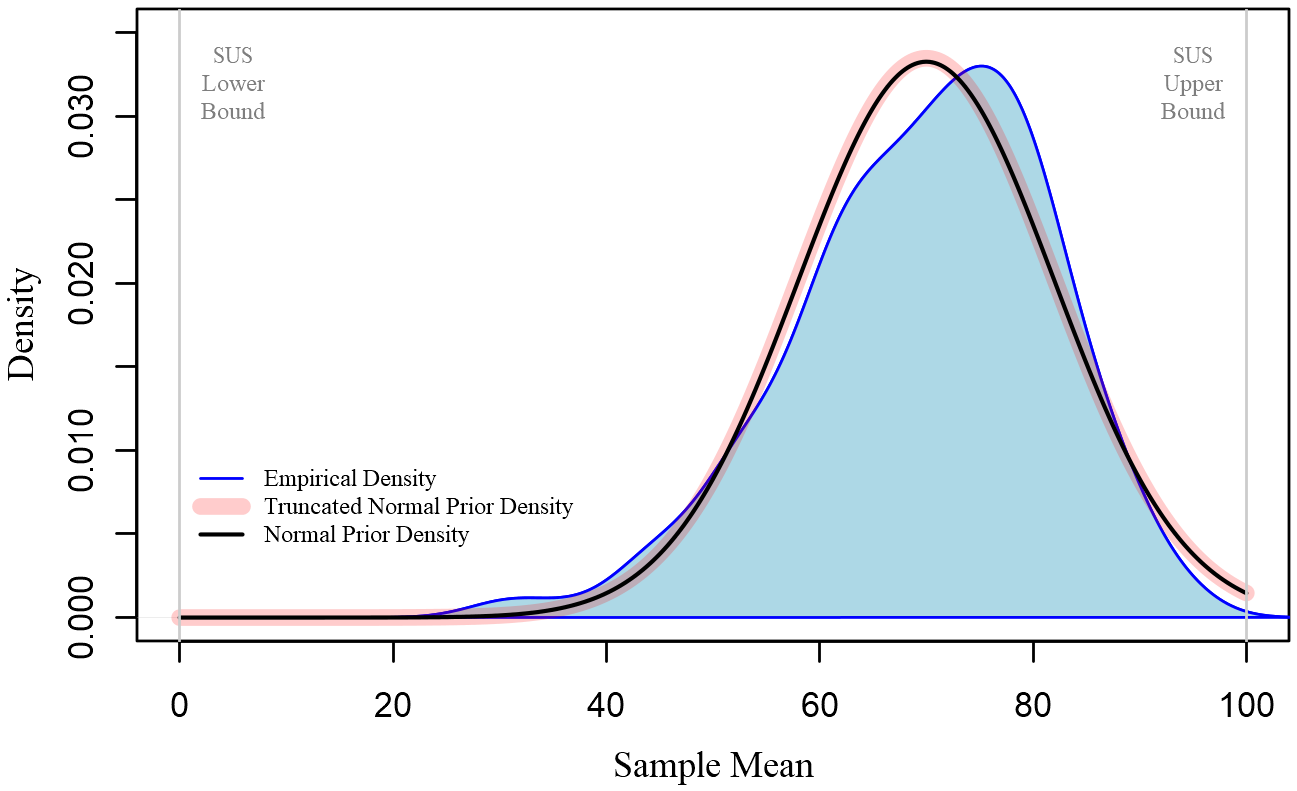}
\caption{Two choices of priors for $\mu$ overlayed on top of the empirical distribution.}
\label{fig:prior}
\end{figure}

Bayesian inference offers an approach that both takes advantage of prior information and properly facilitates inference when $n$ is extremely small by relying on prior beliefs to inform posterior probabilities regarding a parameter of interest. While in some cases this might appear superfluous, historical SUS scores can meaningfully inform inference regarding sample means. Notably, \cite{bangor2008empirical} demonstrated that for 206 usability studies using the SUS, there was never a mean SUS score below 30 or one above 95.  Moreover, their data would suggest a prior distribution (density) regarding the true SUS score mean $\pi(\mu)$ similar to the empirical density denoted by the blue line in Figure~\ref{fig:prior}, created by assuming $\bar{X}$ follows a truncated normal distribution and matching moments. This yields a truncated normal distribution with mean 70 and standard deviation 12 as indicated by the thick pink line in Figure~\ref{fig:prior}. This choice appears to align with the empirical distribution of $\bar{X}$ fairly well, as does a non-truncated normal distribution with mean of 70 and standard deviation of 12.  Thus, as can be seen in Figure~\ref{fig:prior}, although the truncated normal prior distribution is theoretically sound, including truncation in the model to construct a prior density on $\mu$ increases the model's complexity without substantively improving the results. The largest difference between the two exists in the region above 100, and this is only 0.007\% of the data. Lest this appear contradictory to earlier evidence against relying on the CLT, the concern here is on a prior distribution of $\mu$ rather than a sampling distribution of $\bar{X}$. 



The marginal posterior distribution for $\mu$ given the data from any given usability assessment can be estimated by appealing to Bayes theorem as follows: 
 \begin{align}
     \pi(\boldsymbol{\mu},\sigma | \textbf{y}) &\propto \prod_{i,j} f(y_{i,j}|\mu_i,\sigma) \pi(\mu_i)\pi(\sigma) \label{eq:Gen Bayes}.
 \end{align}
 Here $f(y_{i,j})$ is the likelihood function for observation $j$ within test $i$ assuming a truncated Normal distribution. Letting $\pi(\mu_i)$ denote the prior distribution of the true average score of test $i$,  $\pi(\mu_i)$ can be determined from the data in Bangor et al. \citeyearpar{bangor2008empirical}. Similarly, letting $\pi(\sigma)$ denote the prior distribution for the population standard deviation, which is assumed to be  Uniform(0,30) as it is highly unlikely that the standard deviation is greater than 30. The joint posterior distribution for $\mu$ and $\sigma$, $\pi(\boldsymbol{\mu},\sigma | \textbf{y})$, can then be used to find credible intervals for either both parameters, or by integrating out $\sigma$, a marginal posterior distribution for $\mu$.
 
\begin{figure}
\centering
\includegraphics[width=1\linewidth]{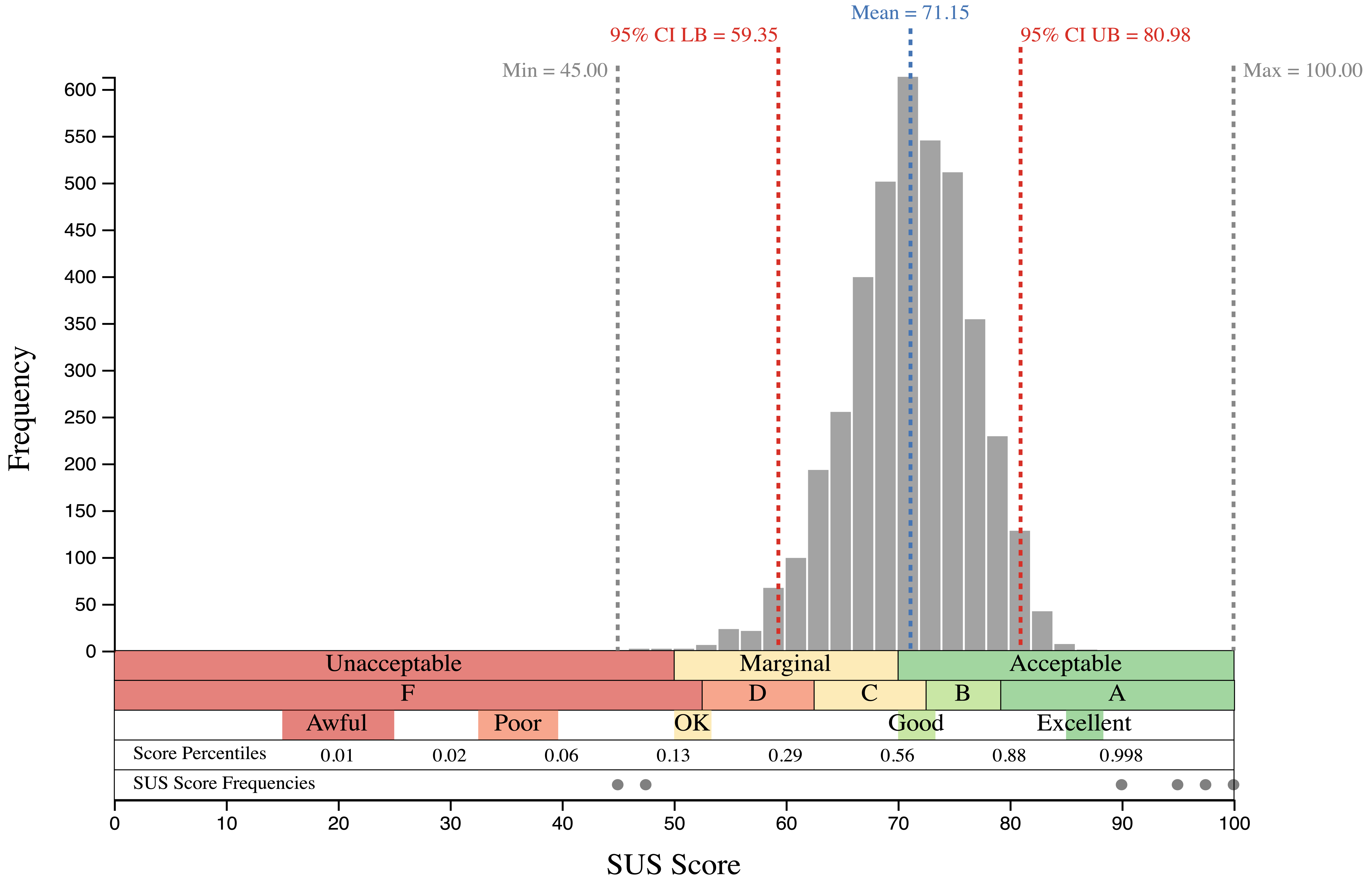}
\caption{95\% empirical Bayesian credible interval for the mean SUS score of the survey highlighted in Figure~\ref{f:SUSscoreBoundViolation1}.}
\label{f:SUSscoreBayes}
\end{figure}

 Typically, \eqref{eq:Gen Bayes} is too complex to evaluate directly without the use of specialized software such as Stan \citep{stan2019}, which relies on Markov Chain Monte Carlo (MCMC) techniques to simulate from the posterior distributions. Practically, an empirical Bayesian approach prevents reporting values for CI or credible interval bounds on $\mu$ that are unrealistic. This is done above by assuming that a truncated normal distribution applies as opposed to a distribution that has support outside of (0,100). Figure~\ref{f:SUSscoreBayes} shows a 95\% empirical Bayesian credible interval for the same SUS study highlighted earlier in  Figure~\ref{f:SUSscoreBoundViolation1}. As demonstrated in what follows, using empirical Bayesian techniques dictate that a practitioner believes that the current study's SUS scores are likely similar to those that have been collected in the past. When sample sizes are extremely small, the mean SUS scores will shift towards the historical norms. It is also important to note that the resulting intervals are not confidence intervals but rather credible intervals that yield a true probability for $\mu$, a useful distinction when communicating results.

\section{Simulation Study}\label{s:simulation_study}

A simulation experiment was conducted to assess the coverage of the expanded BCa bootstrap and \textit{t} distribution CI approaches on very small SUS study results to determine whether 95\% CIs cover the true mean 95\% of the time, and the proportion of CI bounds that fall outside of the SUS score's parameter space.

\subsection{Simulation Methodology and Results}\label{ss.method_results}

\begin{figure}
\centering
\includegraphics[width=1\linewidth]{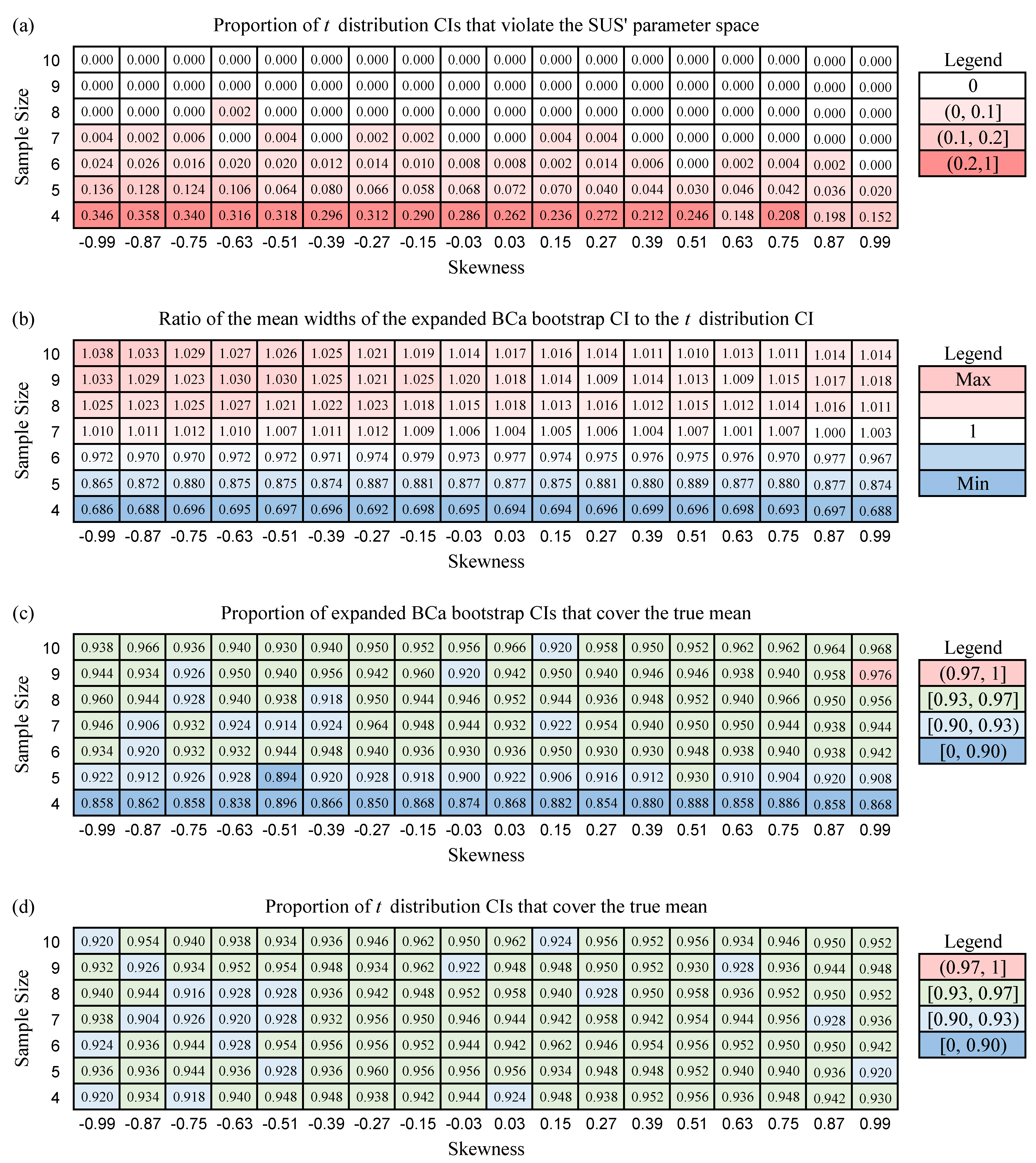}
\caption{Summary of 95\% CI performance for a simulation of 500 samples of size $n \in \{4, 5, \ldots, 10\}$ from a skew normal distribution with a mean of 68, a standard deviation of 20, and a skewness ranging between -0.99 and 0.99. Panel (a) highlights the proportion of the \textit{t} distribution CIs that exceed the bounds of the SUS score's parameter space. Panel (b) shows the ratio of the mean CI widths, where values less than 1 indicate the BCa bootstrap (with Hesterberg's expansion (\citeyear{hesterberg2015teachers})) is narrower. Panels (c) and (d) give the observed coverage for the BCa bootstrap and \textit{t} distribution CIs, respectively.}
\label{f:SimulationResults}
\end{figure}

Small sample SUS studies with 4 to 10 respondents were created using a skew normal distribution \citep{azzalini2005skew} with a mean of 68, a standard deviation of 20, and a skewness ranging between -0.99 and 0.99. These choices mirror those seen in practice \citep{bangor2008empirical, lewis2009factor} while abiding the theoretical bounds for skewness appearing in the skew normal distribution \citep{azzalini1985class}. For each combination of sample size and skewness, 500 sets of scores were generated, along with 95\% CIs for each set using the expanded BCa bootstrap and the \textit{t} distribution techniques. The results of this experiment are summarized in the four panels shown in Figure~\ref{f:SimulationResults}. 

The results illustrated in Panel (a) of Figure~\ref{f:SimulationResults} show that the \textit{t} distribution CIs begin violating the SUS score's parameter space at $n=8$, and the situation gets progressively worse as $n$ decreases. At a skewness of -0.39, the average skewness observed in real-world SUS studies \citep{sauro2016quantifying}, roughly 30\% of the \textit{t} distribution CIs exceed the SUS score's upper bound when $n=4$. As seen earlier, the expanded BCa bootstrap CI's bounds are percentiles of the bootstrap sample drawn with replacement from the observed scores. As a consequence, the expanded BCa bootstrap CI always abides the parameter space. Moreover, as seen in Panel (b), for $n \leq 6$ the expanded BCa bootstrap also tends to produce much narrower intervals. However, Panel (c) demonstrates that this narrowness comes at the expense of coverage, and for $n\leq 5$ the expanded BCa bootstrap CI fails to cover the true mean at an unacceptable rate. The results in Panel (d) shows that the \textit{t} distribution CI's coverage performance is significantly better for extremely small samples, but this is not surprising given its considerably wider intervals - intervals that often violate the SUS score's parameter space. 

In summary, the following decision rules apply: 

\begin{enumerate}
  \item $n \leq 5$: With 5 or fewer respondents, both the expanded BCa bootstrap and the \textit{t} distribution CI have significant shortcomings. Additional information is at a premium, and the empirical Bayesian approach offers a way to harness it, albeit at a cost of having to assume that the current study's mean SUS score will follow similar patterns as previously recorded studies.
  \item $n \in \{6, 7, 8\}$: When compared to the \textit{t} distribution, the expanded BCa bootstrap CI offers acceptable, comparable coverage and narrower or similar widths. It also abides the SUS score's parameter space, and its bounds represent feasible realizations of the true mean SUS score for a sample of size $n$.
  \item $n\geq 9$: With 9 or more respondents, Figure~\ref{f:SimulationResults} suggests the \textit{t} distribution CI abides the parameter space, is slightly narrower than the expanded BCa bootstrap CI, and has good coverage. Based on this finding, one could argue that a practitioner should strive for at least 9 participants when conducting a SUS study, as it would allow for safely constructing traditional CIs using the \textit{t} distribution. Although this logic has merit, being narrower on average does not imply \textit{always} narrower, and it is easy to construct pathological representative SUS score examples where a \textit{t} distribution CI will exceed the SUS score's upper bound. With this in mind, it appears prudent to construct CIs using both the expanded BCa bootstrap and the \textit{t} distribution, and to subsequently pick the one that is both narrower and abiding of the parameter space. Simulation validated this methodology, yielding observed coverage probabilities ranging from a low of 0.92 to a high of 0.96 with an average of 0.943. In short, nominal coverage is preserved.
\end{enumerate}


Although it may seem contrary to the philosophy of Bayesian statistics, for completeness the simulation described above was repeated using credible intervals from the empirical Bayesian methodology outlined in Section~\ref{ss.empiricalBayes}. Echoing a comment made earlier, unlike a traditional confidence interval, a Bayesian credible interval allows for a probabilistic statement to be made directly about the parameter of interest. For example, if $[L, U]$ is a $100(1-\alpha)\%$ credible interval for $\mu$, then there is a $(1-\alpha)$ probability that the population mean lies between $L$ and $U$. To verify that such probabilistic statements are reasonable, the percentage of Bayesian credible intervals that covered $\mu = 68$ over the values of $n$ and skewness used in Figure~\ref{f:SimulationResults} were calculated. These percentages ranged from 92\% to 99\%, which agrees nicely with the intended 95\% chance of the credible intervals containing $\mu$. Nonetheless, this level of agreement should be expected, as the simulation was run with a mean that is close to the mean of the prior distribution, which is 70. If the true mean happened to be far from the mean of the prior distribution, this would not be the case. That said, in the absence of additional SUS scores, the Bayesian approach assumes the system under consideration will most likely have a mean similar to other systems that have been previously studied.

\subsection{Implications for a Practitioner}

While the above simulations appear academic in nature, there are several practical reasons to care about the reporting of the upper confidence bound for SUS study scores. In practice, a usability practitioner can make one of two errors when conducting a SUS study. One could fail to conclude a system is acceptable when it is acceptable, or one could fail to conclude a system is unacceptable when it is unacceptable. The latter appears to be a graver error in the context of assessing usability, as it potentially places an unacceptable system in the hands of the target user base, or it wastes additional resources on further confirmatory studies.

Examining the upper bound of a SUS study score's confidence interval helps the practitioner to avoid this type of error. For example, if the true SUS study score is indeed unacceptable, say 50, one would not want to conclude that the product's usability is acceptable. To test whether earlier results hold in this regard, SUS study scores for $n\in \{4,5,6,7,8,9,10\}$ over a range of skewness were simulated, and the number of times that the upper bounds exceeded 70 for both the \textit{t} distribution and the expanded BCa bootstrap were counted. Overall, 40\% of the \textit{t} distribution confidence intervals contained 70, while the expanded BCa bootstrap intervals contained 70 only 31\% of the time. Furthermore, over the range of the skewness tested, the expanded BCa bootstrap intervals reported fewer errors than the \textit{t} distribution intervals in 62\% of the simulations.

Although the focus here is on the upper bound, if a practitioner's primary concern is with reporting an acceptable system as unacceptable, the lower bound should be emphasized. In such circumstances, simulation testing suggests the \textit{t} distribution and expanded BCa bootstrap intervals perform similarly for $n\in \{6,7,8\}$. Although beyond the scope of this paper, preliminary analysis indicates that truncated \textit{t} distribution intervals are a promising alternative.

\subsection{Validation of Decision Rules against Actual SUS Studies}\label{ss.validation}

\begin{figure*}[!t]
\centering
\subfloat[Stacked dot plot of sample sizes.]{\includegraphics[width=3in]{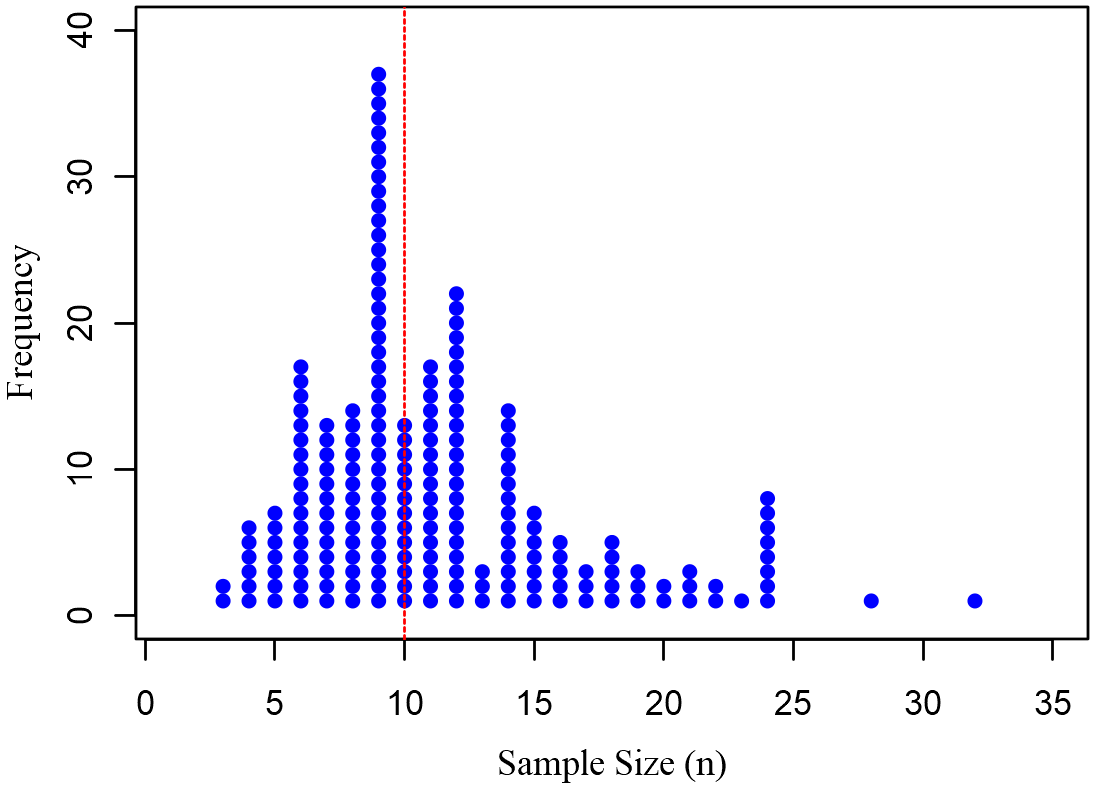}
\label{f.SampleSize}}
\hfil
\subfloat[Density of sample skewness for small sample SUS studies.]{\includegraphics[width=3in]{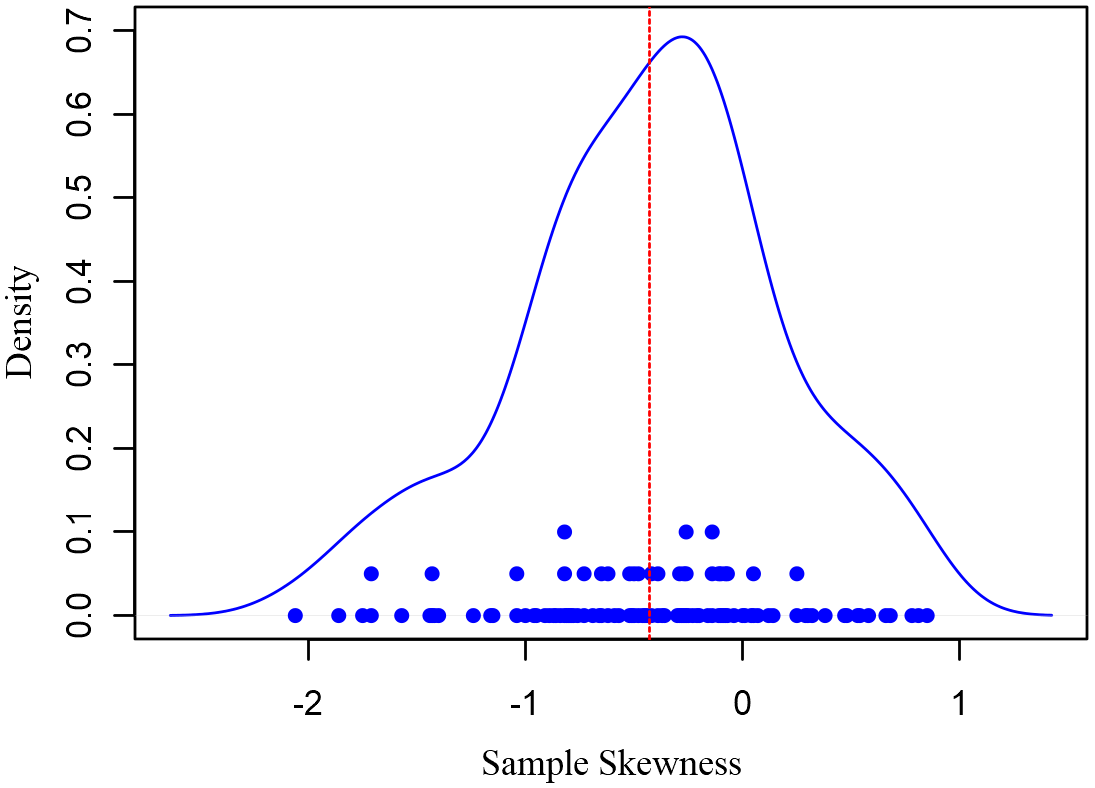}
\label{f:SampleSkewness}}
\centering
\caption{Sample sizes and skewness of the 206 SUS studies in Bangor et al. (\citeyear{bangor2008empirical}).}
\label{f.DescriptiveStats}
\end{figure*}

The effectiveness of the decision rules introduced in the previous section was assessed using a dataset of 206 SUS studies \citep{bangor2008empirical}. As seen in Panel (a) of Figure~\ref{f.DescriptiveStats}, these SUS studies range from a minimum of 3 usability respondents to a maximum of 32 with a median of 10, as indicated by the dashed vertical red line. Additionally, 15 of the studies (7.2\%) had sample sizes of 3 to 5, and 44 (21.4\%) had sample sizes of 6 to 8. Moreover, as Panel (b) of Figure~\ref{f.DescriptiveStats} shows, for the 109 studies with 10 or less respondents, the sample skewness ranged from -2.06 to 0.85 with a mean of -0.43. Taken together, these observations suggest small sample SUS studies are common, and the range of sample skewness simulated earlier is reasonable. Hence, when practitioners find themselves confronted with small sample SUS studies, the procedures outlined above are recommended.

\begin{figure*}[!t]
  \centering
  \subfloat[UCBs using the \textit{t} distribution]{\includegraphics[width=6in]{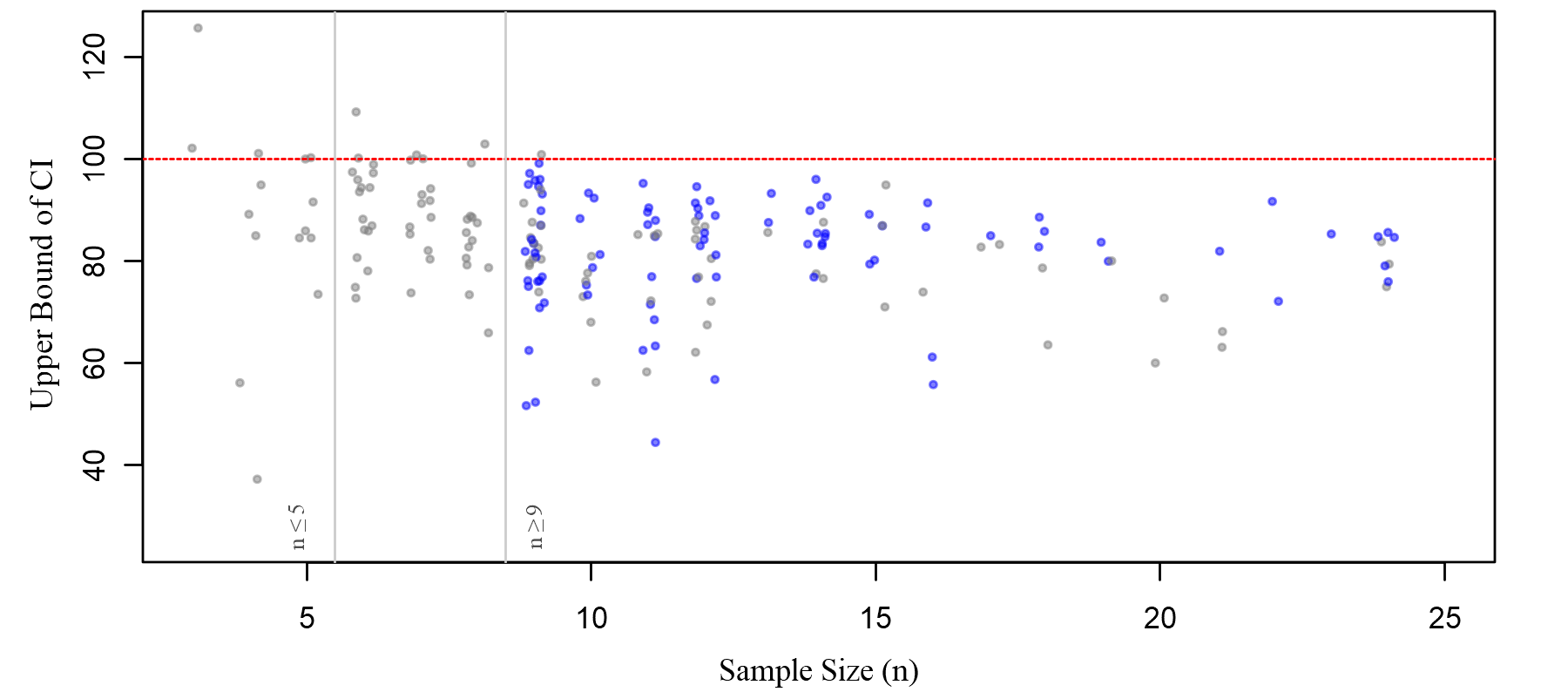}
  \label{fig:UCB2}}
  \hfil
  \subfloat[UCBs using BCa]{\includegraphics[width=6in]{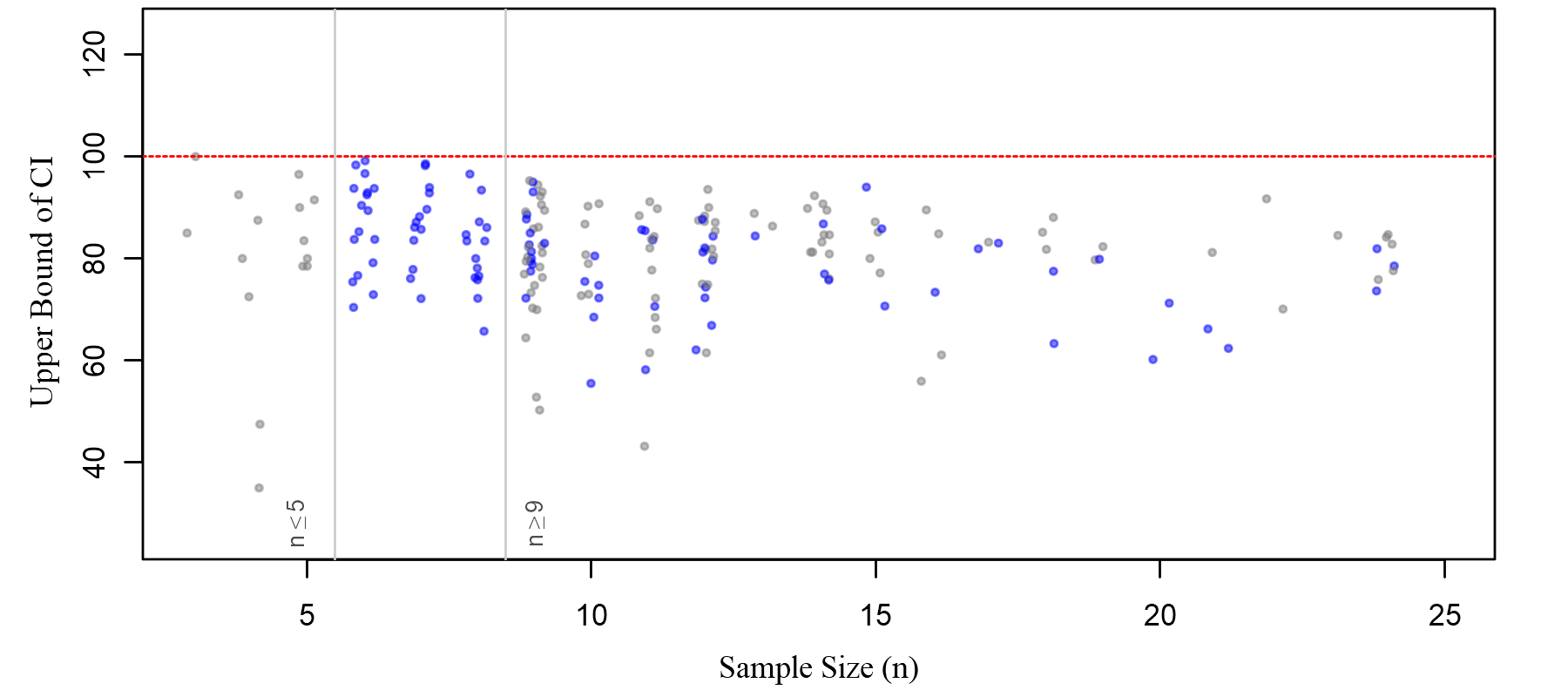}
  \label{fig:BCUCB2}}
  \centering\caption{Plot of sample sizes vs. upper confidence bounds for the data of \cite{bangor2008empirical}, where the sample sizes have been jittered for readability. For each of the 206 SUS studies represented in these plots, a blue colored point denotes that the associated CI construction method is preferred based on the decision rules outlined in the previous section. As seen in Panel (a), it is common for the \textit{t} distribution CIs' upper confidence bounds to exceed the parameter space for $n \leq 8$. Additionally, there are two studies with $n>25$; in both cases, the \textit{t} distribution CI is preferred.}
  \label{f.new_methods_performance2}
  \end{figure*}

To investigate Bangor et al.'s (\citeyear{bangor2008empirical}) dataset further, 95\% CIs for the mean of each SUS study were constructed using the \textit{t} distribution, expanded BCa bootstrap, and empirical Bayesian methods. Panels (a) and (b) of Figure \ref{f.new_methods_performance2} show the upper confidence bounds (UCBs) of these CIs for the \textit{t} distribution and the expanded BCa bootstrap methods, respectively. Additionally, within each plot, the blue colored points denote that the associated CI construction method is preferred based on the decision rules outlined in the previous section. 

Applying the first decision rule, for $n \in \{3,4,5\}$ neither the \textit{t} distribution nor the expanded BCa bootstrap methods should be trusted to generate 95\% CIs that abide the SUS score's parameter space and attain the nominal coverage. Accordingly, none of the points for $n \leq 5$ are colored blue in either of Figure \ref{f.new_methods_performance2}'s panels. Moving to the second decision rule, for samples with 6 to 8 respondents Panel (a) shows that the UCBs of the \textit{t} distribution CIs often exceed the SUS score's maximum value of 100, while Panel (b) highlights that the UCBs of the expanded BCa bootstrap CIs abide the parameter space. In accordance with the second decision rule, all of the points in Panel (b) for $n \in \{6,7,8\}$ are blue. Finally, for samples with 9 or more respondents, the third decision rule implies that if the \textit{t} distribution CI abides the parameter space and is narrower than the expanded BCa bootstrap CI, its representative point in Panel (a) will be blue. Otherwise, the expanded BCa bootstrap CI is the better option, and the associated point in Panel (b) will be blue. Notably, the expanded BCa bootstrap CI is the preferred option in 53 of the 147 SUS studies (36\%) with $n \geq 9$, including several of the studies with the largest number of respondents. Clearly, the expanded BCa bootstrap method is not simply a small sample solution. 

\begin{figure}[!t]
\centering
\includegraphics[width=1\linewidth]{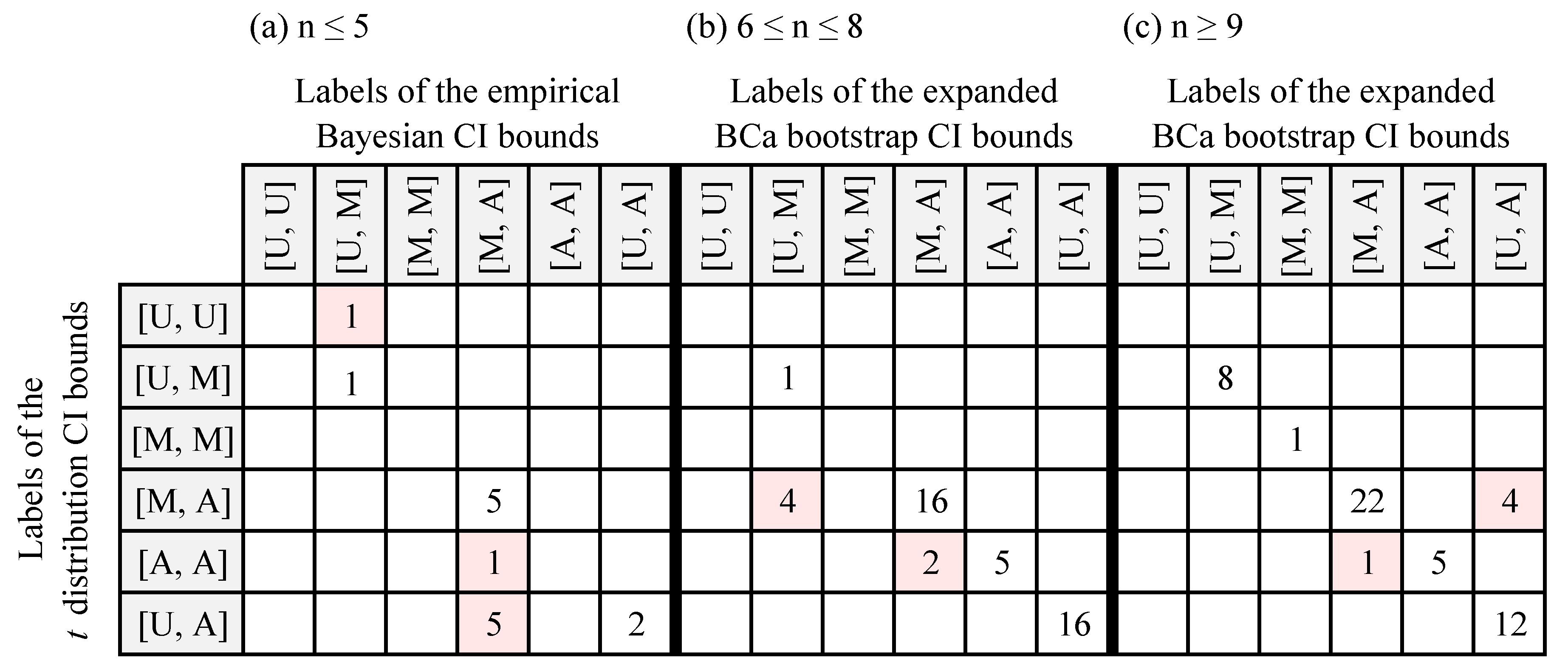}
\caption{Comparison of the labels corresponding to the 95\% CI bounds of the gray points in Panel (a) of Figure \ref{f.new_methods_performance2}, where U = unacceptable, M = marginal, and A = acceptable. Specifically, for $n \leq 5$, the \textit{t} distribution CI is compared to the empirical Bayesian CI, and for $n \geq 6$, it is compared to the expanded BCa bootstrap CI. The numbers in the cells represent counts, and pink colored cells highlight SUS studies where the acceptability labels corresponding to the CIs' bounds disagree.}
\label{fig:labels}
\end{figure}

Figure \ref{fig:labels} provides a tabular summary of the practical consequences of utilizing the empirical Bayesian or expanded BCa bootstrap methods in lieu of the commonly used \textit{t} distribution. In particular, it displays the agreement between the acceptability labels corresponding to the bounds of the \textit{t} distribution CI and the alternative CI suggested by the decision rules, where the numbers in the cells represent counts. For example, when $n \leq 5$ the first decision rule suggests the empirical Bayesian method should be applied. In Bangor et al.'s (\citeyear{bangor2008empirical}) dataset, there are 15 SUS studies that fall into this category. After applying both the \textit{t} distribution and empirical Bayesian methods, there are 7 SUS studies where the acceptability labels disagree. As seen in Panel (a), in 5 of these studies the \textit{t} distribution CI suggests the system's usability ranges from unacceptable to acceptable, while the empirical Bayesian CI sees it as marginal to acceptable. Although the underlying reason for the extremely small number of respondents is unknown, it is reasonable to assume that one or more of the practical limitations mentioned earlier are present. Applying the more appropriate empirical Bayesian method has returned tighter results and sharpened the conclusiveness of the studies. For the remaining two studies in Panel (a), the opposite is true, as the empirical Bayesian method suggests the plausible range of acceptability labels should be loosened to include marginal.     

Moving to the second decision rule, in Panel (b) of Figure \ref{fig:labels} the agreement between the acceptability labels improves, as only 6 of 44 SUS studies (13.6\%) disagree. In all 6 cases, the expanded BCa bootstrap method returned a more pessimistic usability result than the \textit{t} distribution, and in 3 of the 6 the upper bound of the \textit{t} distribution CI exceeded 100. Finally, after applying the third decision rule, there are 53 SUS studies where the expanded BCa bootstrap CI is preferred over the \textit{t} distribution CI. Among these 53 studies, Panel (c) shows only 5 (9.4\%) disagree, and once again, each disagreement is due to the more pessimistic usability result of the expanded BCa bootstrap CI.

\section{Discussion and Custom Application}\label{s:app}

While oftentimes the boundaries of the confidence interval shift only slightly, Figure \ref{fig:labels} highlights that such shifts may affect the usability labels assigned to system usability constructs. Given that practitioners make decisions based on these labels, accurately calculating the boundaries of the confidence interval has practical significance. To this end, the empirical Bayesian or expanded BCa bootstrap methods should be used in practice; however, the statistical acumen necessary for a usability practitioner to employ them presents a potential barrier. With this in mind, the authors developed an intuitive, freely accessible online application that automates the calculation of and the decision rules for these alternative CIs (available at \url{http://sus.dse-apps.com}). The user interface accepts and processes SUS data, provides a recommendation for which method(s) to use, and creates effective visualizations to communicate results to both technical and non-technical clients.  This application complements a recently introduced free mobile device application called \textit{SUSapp} \citep{xiong2020} that helps practitioners administer the SUS and collect data for later analysis.

While the full code base is available under an open source license, the application removes the tasks of configuring and running the code which can be complicated by dependencies (e.g., the need to work with Stan \citep{stan2019} for Bayesian computation). Furthermore, correctly interpreting the direct code output can be difficult without prior knowledge of each method.  

\begin{figure}[htbp]
\centering
\includegraphics[width=15cm]{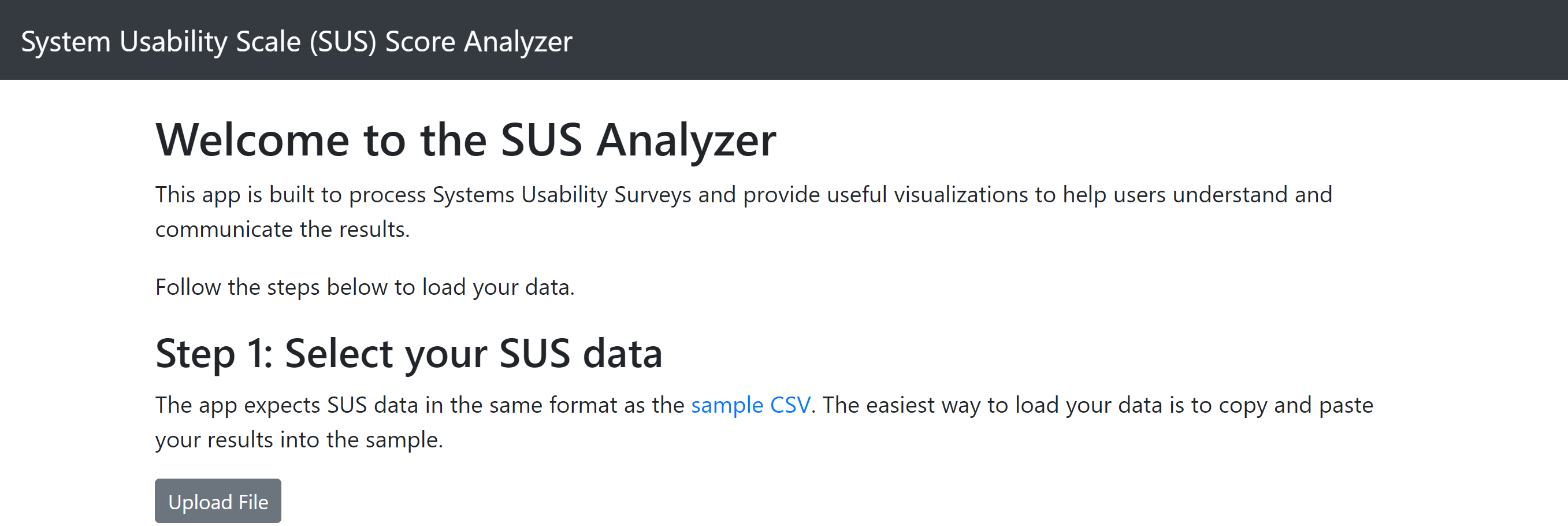}
\caption{The System Usability Scale (SUS) Analyzer application splash page.}
\label{fig:sc_main}
\end{figure}

\begin{figure}[htbp]
\centering
\includegraphics[width=15cm]{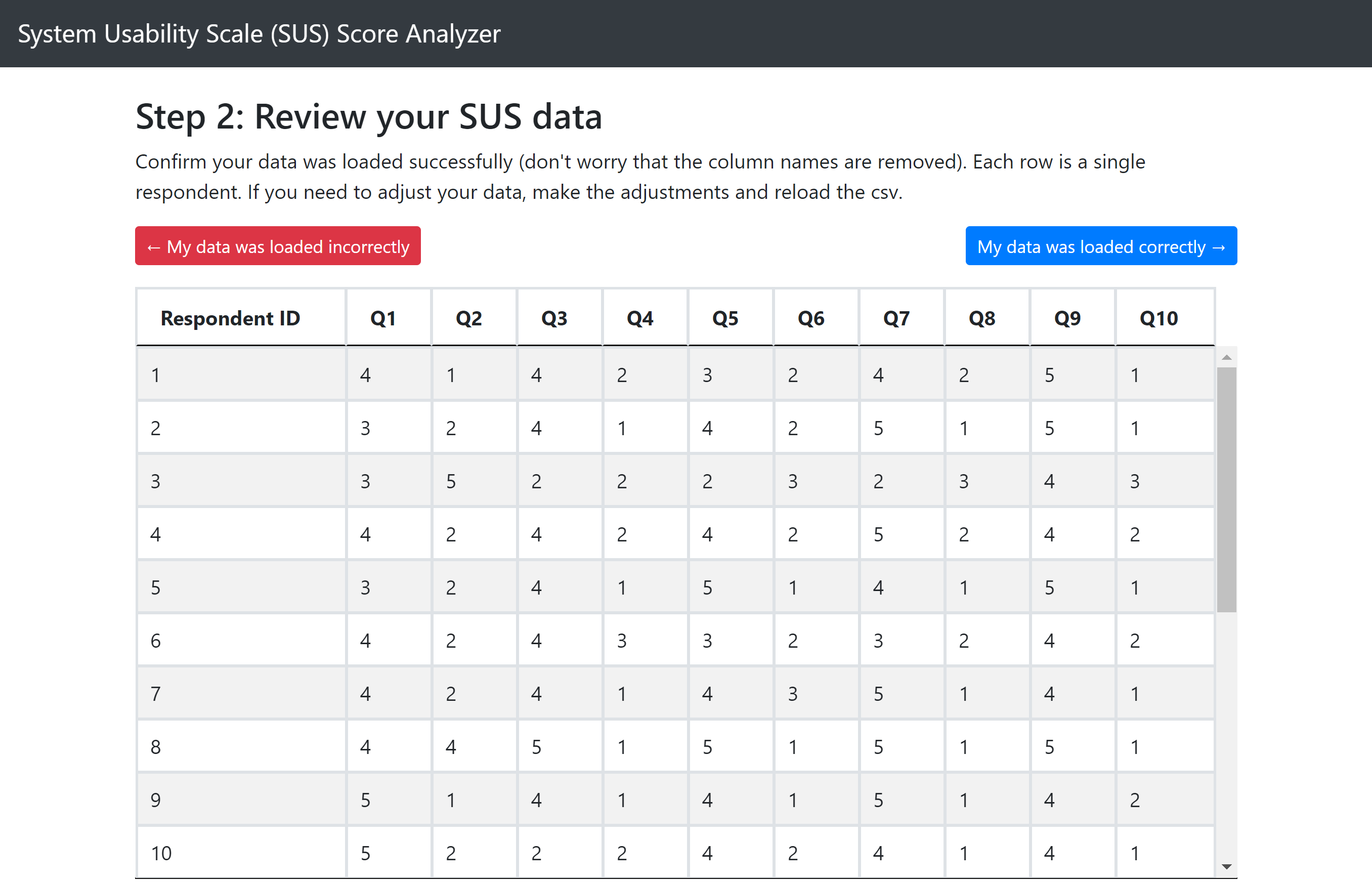}
\caption{Users are able to verify their data were loaded correctly before proceeding to the recommendations and visualizations.}
\label{fig:sc_validate}
\end{figure}

\begin{figure}[htbp]
\centering
\includegraphics[width=15cm]{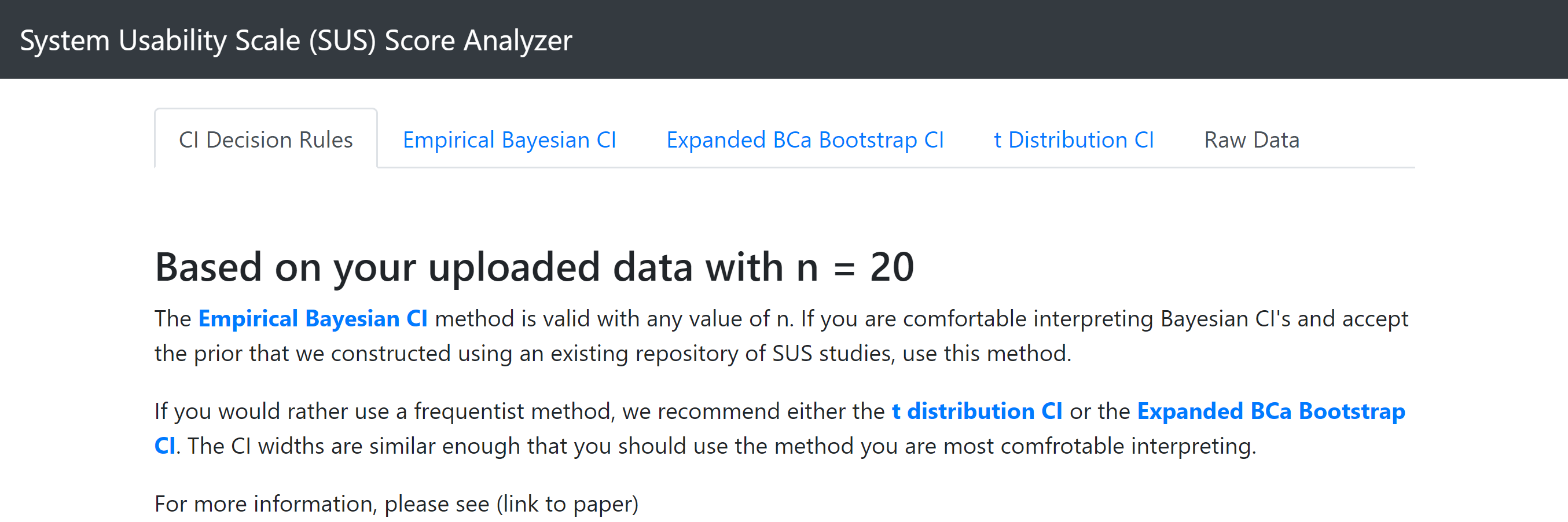}
\caption{The app recommends which method(s) to use based on the decision rules presented in this paper.}
\label{fig:sc_reco}
\end{figure}

The application accepts data in CSV format where each row represents an individual's response to the SUS questionnaire. As seen in Figure \ref{fig:sc_main}, a sample data set showing the required formatting is available on the welcome page via the blue hyperlink. After loading data using the ``Upload File" button, a facility is provided for users to view and confirm that data were loaded correctly (see Figure \ref{fig:sc_validate}). Once complete, users select the appropriate button to either edit input data or submit their data for the decision rules to be applied as described in this paper (see Figure \ref{fig:sc_reco}).   Recommended methods that adhere to these decision rules appear as blue tabs.  If a particular method is not recommended (e.g., an option of using the \textit{t} distribution method with a sample size lower than 5), the tab will display as grey, and a user will not be able to select it. 

\begin{figure}[htbp]
\centering
\includegraphics[width=15cm]{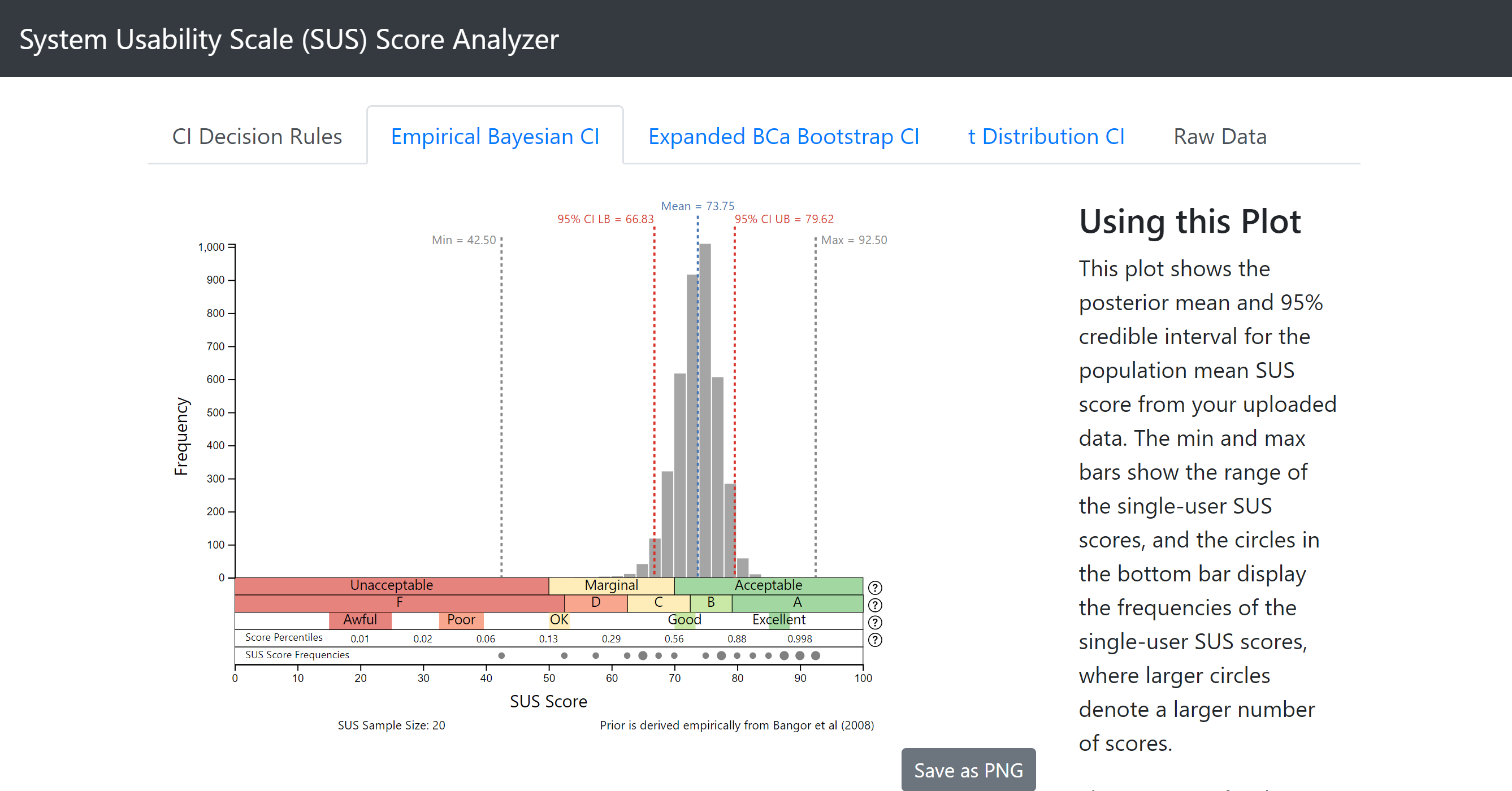}
\caption{Results of empirical Bayesian analysis on SUS data showing a 95\% credible interval.}
\label{fig:sc_bayes}
\end{figure}

A user can move between the recommended tabs to see visualizations for Bayesian, expanded BCa bootstrap, and $t$ distribution frequency plots, resulting means and confidence intervals, and appropriate mapping to the four common scales mentioned in Section~\ref{s.introduction}. All visualizations can be exported in high resolution images using the ``Save as PNG" button at the bottom of each plot (see Figure \ref{fig:sc_bayes}).

While there are many additional features that could be included in this application to meet the needs of specific user bases, the version available is intended as a general purpose tool for usability practitioners. Because the source code is available under MIT open source license, practitioners needing to modify the application, its underlying analysis options, decision rules, or output visualizations can do so to add features or modify any of the methods to accommodate specific needs.

\section{Conclusions}\label{s:conclusions}

The effectiveness of the SUS for assessing usability of systems is well-established among usability practitioners. When a sufficient number of survey respondents are available to leverage the central limit theorem for analyzing and reporting SUS study uncertainties, current practices invoking either a normal distribution or $t$ distribution for constructing confidence intervals are sound.  However, when only a small number of users are surveyed, as in cases in which the desired user pool is not available or affordable, reliance on the central limit theorem yields confidence intervals that suffer from parameter bound violations and interval widths that confound mappings to adjective and other constructed scales that organizations rely upon for decision making. These shortcomings are especially pronounced when the underlying SUS score data is skewed, as it is in many instances. Using actual SUS data made available for this study, the $t$ distribution's specific inadequacies are illustrated.

Unfortunately, when the sample size is small there is not a single tool that a user can apply in all situations. This paper helps to remedy this by introducing two attractive alternatives that improve the accuracy and reporting of SUS study results when practitioners are faced with very small ($n \leq 10$) and extremely small ($n \leq 5$) samples, namely the expanded BCa bootstrap and an empirical Bayesian approach. These alternative approaches facilitate three novel decision rules when constructing confidence or credible intervals on small sample SUS study means. Additionally, a freely accessible, online application for the usability practitioner to implement these decision rules and produce effective visualizations is developed and presented for general use.

It is important to note that for the empirical Bayesian approach the interpretation of the resultant interval is not the same as parametric or non-parametric confidence intervals.  In fact, under this paradigm the definition of probability itself is different.  Specifically, Bayesian inference relies on subjective probability, meaning it is a measure of belief rather than the long-run frequency.  As such, the posterior distribution and associated statistics represent a belief in the population mean, and probabilities can be directly calculated from it.  So, if the interval for a mean SUS score is, for example, (45, 75), then it would be correct to conclude that there is a 95\% probability that the population mean is between 45 and 75.  While this interpretation is only available when using the empirical Bayesian construction and not the expanded BCa bootstrap technique, it does provide a much more intuitive and natural interpretation regarding uncertainty than a confidence interval, offering a small but significant advantage when communicating SUS results to a non-technical client.

While this paper expands the current options for practitioners in reporting SUS scores, there are several questions yet to be explored. Particularly, the current empirical Bayesian approach does not account for the differing types of systems whose usability is assessed using the SUS. For example, a software program may have important inherent differences than a cellphone. To account for system-to-system differences, a hierarchical Bayesian model could potentially be useful. However, recognizing that more nuanced models such as this require additional data to develop, practitioners should be encouraged to share their SUS data to the maximum extent possible for the benefit of the broader usability community.  In this spirit of continuous improvement, the authors welcome any feedback or suggestions for improving the online application. Ultimately, our hope is that usability practitioners will gain value from its use in this setting or in other topical areas which exhibit the type of problem characteristics addressed in this paper.

\section*{Data Availability}
\addcontentsline{toc}{section}{Data_Availability}
The data that support the findings of this study were obtained from the corresponding author of Bangor et al. (2008). Restrictions apply to the availability of these data, and it is unavailable.


\bibliography{SUSreferences}

\end{document}